\documentclass[twocolumn,preprintnumbers,superscriptaddress]{revtex4}
\usepackage{graphicx}
\usepackage{dcolumn}
\usepackage{bm}
\usepackage{color}
\usepackage{amsmath}
\usepackage{amssymb}
\usepackage{amsfonts}
\usepackage{esint}
\usepackage{times}
\usepackage{url}
\usepackage{setspace}

\usepackage[colorlinks,linkcolor=blue,anchorcolor=blue,citecolor=blue]{hyperref}

\begin{document}
\title{Switchable next-nearest-neighbor coupling for controlled two-qubit operations}
\author{Peng Zhao}\email{shangniguo@sina.com}
\affiliation{National Laboratory of Solid State Microstructures, School of Physics, Nanjing University, Nanjing 210093, China}
\author{Peng Xu}\email{pengxu@njupt.edu.cn}
\affiliation{Institute of Quantum Information and Technology, Nanjing University of Posts and Telecommunications, Nanjing, Jiangsu 210003, China}
\affiliation{National Laboratory of Solid State Microstructures, School of Physics, Nanjing University, Nanjing 210093, China}
\affiliation{State Key Laboratory of Quantum Optics and Devices, Shanxi University, Taiyuan, 030006, China}

\author{Dong Lan}\affiliation{National Laboratory of Solid State Microstructures, School of Physics, Nanjing University, Nanjing 210093, China}
\author{Xinsheng Tan}\email{meisen0103@163.com}\affiliation{National Laboratory of Solid State Microstructures, School of Physics, Nanjing University, Nanjing 210093, China}
\author{Haifeng Yu}\affiliation{National Laboratory of Solid State Microstructures, School of Physics, Nanjing University, Nanjing 210093, China}
\author{Yang Yu}
\affiliation{National Laboratory of Solid State Microstructures, School of Physics, Nanjing University, Nanjing 210093, China}

\date{\today}

\begin{abstract}
In a superconducting quantum processor with nearest neighbor coupling, the dispersive
interaction between adjacent qubits can result in an effective
next-nearest-neighbor coupling whose strength depends on the
state of the intermediary qubit. Here, we theoretically explore the possibility of engineering
this next-nearest-neighbor coupling to implement controlled two-qubit operations
where the intermediary qubit controls the operation on the next-nearest neighbor pair
of qubits. In particular, in a system comprising two types of superconducting qubits
with anharmonicities of opposite-sign arranged in an -A-B-A- pattern, where the
unwanted static ZZ coupling between adjacent qubits could be heavily
suppressed, a switchable coupling between the next-nearest-neighbor qubits can be
achieved via the intermediary qubit, the qubit state of which functions as an on/off switch for this coupling.
Therefore, depending on the adopted activating scheme, various controlled two-qubit operations
such as controlled-iSWAP gate can be realized, potentially enabling circuit depth
reductions as to a standard decomposition approach for implementing generic
quantum algorithms.
\end{abstract}

\maketitle

\section{Introduction}

Implementing a gate-based quantum processor relies on arrays of qubits coupled
together, and in a quantum processor with superconducting circuits, nearest
neighbor (NN) coupling via a linear circuit (e.g., capacitor) provides a native
architecture to satisfy this requirement \cite{R1}. But in practice, aside from the dedicated
designed NN coupling, next-nearest-neighbor (NNN) coupling can also present in
the superconducting quantum processor via several different mechanisms \cite{R2,R3}, such as
unintended static directly or indirectly capacitive/inductive coupling described as
a two-body interaction between NNN qubits via an effective reactance, and effective
quantum coupling between NNN qubits resulting from the dispersive interaction
between NN qubits. In general, these NNN couplings are considered as unwanted
spurious interaction between qubits, leading to gate infidelities, thus various
approaches have been proposed to suppress these spurious interactions \cite{R4,R5}.

However, at the same time, the NNN coupling could also be utilized as a dedicated
channel for implementing non-trivial tasks \cite{R5}. In particular, the effective NNN coupling
mediated by the intermediary qubit could be explored to realize a native three-qubit
gate without resorting to the decomposition approach that involves a series of
single- and two-qubit gates, thus reducing circuit depth for quantum algorithms
and making them potentially attractive for NISQ application \cite{R6}. This is enabled by the fact that this
NNN coupling is essentially a native three-body interaction \cite{R7}, acting as a
natural resource for implementing three-qubit gate operations. However, since this
effective coupling is enabled by a second order process, its strength has a magnitude similar to the
residual two-body interaction between adjacent qubits, such as ZZ coupling \cite{R8,R9}, which imposes
a limiting factor on the performance of the native three-qubit gate. We note that various theoretical and
experimental studies have previously explored this three-body interaction, but in the situation where the
interaction is commonly used as a two-body interaction for two-qubit gate operations by
setting the intermediary qubit (treated as a bus coupler) in its ground state \cite{R7,R10,R11,R12} (see also Appendix A).

\begin{figure}[tbp]
\begin{center}
\includegraphics[width=8cm,height=4cm]{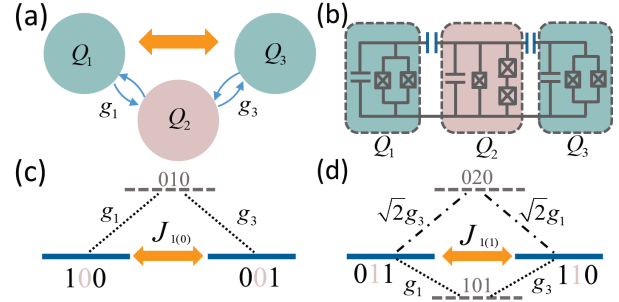}
\end{center}
\caption{(a) Sketch of an -A-B-A-type three-qubit system with NN coupling. Dispersive interaction
between adjacent qubits (denoted as round arrows) can result in an effective
NNN coupling (denoted as double-headed arrows). (b) Circuit diagram of a chain of
three superconducting qubits capacitively coupled to each other, where the $Q_{1(3)}$
and $Q_{2}$ are transmon qubit \cite{R21} and C-shunted flux qubit \cite{R22,R23,R24},
which can be modeled as weak anharmonic oscillators with negative and positive
anharmonicity, respectively. (c) For the intermediary qubit $Q_{2}$
in $|0\rangle$ state, the NNN exchange interaction with strength $J_{1(\textbf{0})}$
is enabled by a single path (denoted as dashed lines) involving
the intermediate state $|010\rangle$. (d) For $Q_{2}$ in
$|1\rangle$ state, the NNN interaction with strength $J_{1(\textbf{1})}$
is enabled by two paths (denoted as dashed lines and
dash-dotted lines, respectively), and each involves an intermediate state, i.e.,
$|101\rangle$ or $|020\rangle$.}
\end{figure}

In this work, we theoretically explore the possibility of engineering the (intermediary)
qubit-mediated NNN coupling in a scalable superconducting quantum processor to implement controlled
two-qubit operations where the intermediary qubit controls the operation on the NNN qubits.
We demonstrate that, by coupling two types of superconducting qubits with anharmonicities of opposite signs
arranged in an -A-B-A- pattern, on the one hand, the unwanted ZZ coupling between
adjacent qubits can be heavily suppressed \cite{R13,R14}, thus breaking the limitation on the performance of
potentially implemented native three-qubit gates, on the other hand, a switchable coupling
between NNN qubits can be realized \cite{R15}, where the intermediary qubit state functions
as an on/off switch for this NNN coupling. Thus, depending on the activating scheme, this
switchable NNN coupling could be used to realize various controlled two-qubit operations \cite{R16},
and a case study shows that controlled-iSWAP (C-iSWAP) gate \cite{R17,R18} with intrinsic gate fidelity
(excluding the decoherence error) in excess of $99.9\,\%$ can be achieved in $50$ ns. We further show that
compared with the decomposition methods based on single- and two-qubit gates, textbook three-qubit gates
such as Toffoli gate (CCNOT) and Controlled-Controlled-Z gate (CCZ), which are widely used in various
quantum circuits \cite{R19}, can be constructed efficiently via the native C-iSWAP gate. Moreover, the switchable
NNN coupling can also be employed for efficiently generating a continuous set of controlled two-qubit gates
for quantum chemistry \cite{R20}.

\section{switchable NNN exchange coupling}

\begin{figure}[tbp]
\begin{center}
\includegraphics[width=8cm,height=9cm]{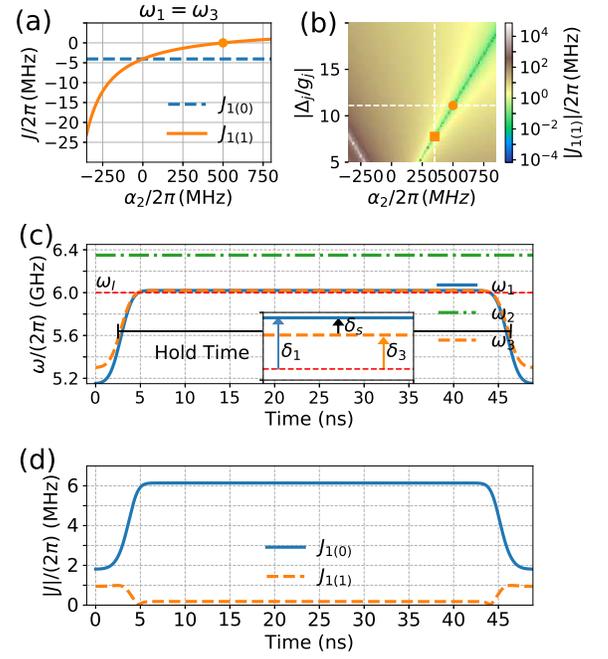}
\end{center}
\caption{(a) Calculated strength of the effective NNN interactions $J_{1(\textbf{1})}$
and $J_{1(\textbf{0})}$ versus $Q_{2}$'s anharmonicity $\alpha_{2}$ with NN coupling
strength $g_{j}/2\pi=45\,\rm MHz$, qubit detuning $\Delta_{j}/2\pi=-500\,\rm MHz$, and
qubit anharmonicity $\alpha_{j}/2\pi=-350\,\rm MHz$. The orange dot indicates the working
point where the NNN couping is turned off for $Q_{2}$ in $|1\rangle$ state. (b) Calculated
values of $J_{1(\textbf{1})}$ versus $\alpha_{2}$ and $\Delta_{j}$ in the unit of $g_{j}$.
The anharmonicity $\alpha_{j}$ and NN coupling strength $g_{j}$ take the same values as
in (a). The horizontal cut denotes the result plotted in (a), and the intersection (orange
square) of the vertical cuts at $\alpha_{2}/2\pi=\,\rm 350 MHz$ and the blue strip gives
the ideal optimal working point for native C-iSWAP gate operations. (c) Typical control
pulse for realizing C-iSWAP gate, where the frequency of $Q_{2}$ is fixed, and the
frequencies of $Q_{1}$ and $Q_{3}$ vary from the idle frequency point $\omega_{i}$ to the
interaction frequency point $\omega_{I}$ and then come back. The inset
highlights the frequency offset $\delta_{1(3)}$ with respect to the ideal interaction
point (horizontal red dashed line), where $\delta_{s}=\delta_{1}-\delta_{3}$ denotes
a small frequency overshoot. (d) Calculated values of the analytical strength of the effective exchange interactions
$J_{1(\textbf{0})}$ and $J_{1(\textbf{1})}$ as a function of time with the typical control pulse (c)
applied to the three-qubit system. }
\end{figure}

To start let us consider an A-B-A type system comprising three superconducting
qubits (labeled as $Q_{1,2,3}$, where $Q_{1(3)}$ (A-type qubit) denotes the transmon qubit
with negative anharmonicity, and $Q_{2}$ (B-type qubit) represents the capacitively shunted
flux qubit with positive anharmonicity \cite{R21,R22,R23,R24}) capacitively coupled to each
other, as depicted in Fig.$\,$1, which can be modeled by a chain of three weakly anharmonic
oscillators with NN coupling, described by (hereinafter $\hbar =1$)
\begin{eqnarray}
\begin{aligned}
H=&\sum_{l}\left[\tilde{\omega}_{l}q_{l}^{\dagger}q_{l}+\frac{\alpha_{l}}{2}q_{l}^{\dagger}q_{l}^{\dagger}q_{l}q_{l}\right]
+\sum_{j}g_{j}(q_{j}^{\dagger}q_{2}+q_{j}q_{2}^{\dagger}),
\end{aligned}
\end{eqnarray}
where the subscript $l={1,2,3}$ labels qubit $Q_{l}$ with anharmonicity
$\alpha_{l}$ and bare qubit frequency $\tilde{\omega}_{l}$, $q_{l}\,(q_{l}^{\dagger})$ is
the associated annihilation (creation) operator, and $g_{j}$ ($j=1,3$) denotes the
strength of the NN coupling between $Q_{j}$ and $Q_{2}$. In the following
discussion, we focus on this -A-B-A-type system for implementing switchable
NNN coupling and controlled two-qubit gate operations, leaving the extension to the -B-A-B-type
system and even the two-dimensional array to the discussion.

We now consider that the system operates in the dispersive regime where the
qubit frequency detuning ($\Delta_{j}=\tilde{\omega}_{j}-\tilde{\omega}_{2}$)
is far larger than the NN coupling strength, i.e., $|\Delta_{j}|\gg g_{j}$.
Thus, by using the Schrieffer-Wolff transformation \cite{R25,R26} which removes the NN
coupling in the Hamiltonian given in Eq.$\,$(1), one can obtain an effective
block-diagonal Hamiltonian for the full system (see Appendix B for a full derivation).
Truncated to the qubit levels, the effective Hamiltonian has the following form (see
Appendix B for a full derivation)

\begin{eqnarray}
\begin{aligned}
H_{\rm eff}=&\omega_{1}\frac{ZII}{2}+\omega_{2}\frac{IZI}{2}+\omega_{3}\frac{IIZ}{2}
+\zeta_{1}\frac{ZZI}{2}+\zeta_{3}\frac{IZZ}{2}
\\&+J_{Z}\frac{XZX+YZY}{2}+J_{I}\frac{XIX+YIY}{2}
\\&+\zeta_{Z}\frac{ZZZ}{2}+\zeta_{I}\frac{ZIZ}{2},
\end{aligned}
\end{eqnarray}
where $\rm(X,Y,Z,I)$ represent the Pauli operator and identity operators, and
the order indexes the qubit number, $\omega_{l}$ and $\zeta_{j}$ denote
dressed qubit frequency of $Q_{l}$ and the strength of the ZZ coupling between
$Q_{j}$ and $Q_{2}$, respectively. The last four terms represent the effective
interaction between NNN qubits ($Q_{1}$ and $Q_{3}$).

In Eq.$\,$(2), the terms XZX+YZY and XIX+YIY result in a net virtual exchange
interaction between $Q_{1}$ and $Q_{3}$, and the value of its net strength
depends on the state of $Q_{2}$. From the view of the second-order perturbation
theory, the physics behind this feature is that the virtual exchange interaction
results from different contributions, depending on the state of $Q_{2}$.
As shown in Fig.$\,$1(c), in an -A-B-A-type three-qubit system where the frequency of $Q_{2}$ satisfies
$\tilde{\omega}_{2}>\tilde{\omega}_{j}$, the effective interaction between $|100\rangle$
and $|001\rangle$ with strength $J_{1(\textbf{0})}=J_{I}-J_{Z}=
g_{1}g_{3}(\Delta_{1}+\Delta_{3})/(2\Delta_{1}\Delta_{3})$ is enabled by the
path given as $|100\rangle\rightarrow |010\rangle\rightarrow |001\rangle$. However,
as shown in Fig.$\,$1(d), for the effective interaction $|110\rangle\leftrightarrow|011\rangle$ with
strength $J_{1(\textbf{1})}=J_{I}+J_{Z}$ given as
\begin{eqnarray}
\begin{aligned}
J_{1(\textbf{1})}=\frac{g_{1}g_{3}}{2}\left[\frac{\Delta_{1}+\alpha_{2}}{\Delta_{1}
(\Delta_{1}-\alpha_{2})}+\frac{\Delta_{3}+\alpha_{2}}{\Delta_{3}(\Delta_{3}-\alpha_{2})}\right],
\end{aligned}
\end{eqnarray}
there are two paths given as $|110\rangle\rightarrow |020\rangle\,(|101\rangle)\rightarrow |011\rangle$,
and since $\tilde{\omega}_{2}>\tilde{\omega}_{j}$, the two paths contribute with
strength of opposite-sign, enabling competition between the positive and the
negative contributions. Thus, one may reasonably expect that by engineering
the system parameters, the strength of the dispersive interactions $J_{1(\textbf{1})}$
can take a value of $0$ when the two competitive contributions destructively interferes,
while $J_{1(\textbf{0})}$ is intact.

For our proposed -A-B-A-type system shown in Fig.$\,$1(c),
the B-type qubit (C-shunt flux qubit) has an positive anharmonicity, i.e., $\alpha_{2}>0$,
and the qubit detuning $\Delta_{j}<0$. When $\alpha_{2}=-\Delta_{1}=-\Delta_{3}$, the two
competitive contributions in Eq.$\,$(3) yield zero net coupling strength $J_{1(\textbf{1})}=0$.
Thus, a switchable NNN coupling can be realized. In addition, according to Eq.$\,$(3), a similar
result can also be obtained for a -B-A-B-type three-qubit system, where the anharmonicity
of A-type qubit (transmon qubit) takes a negative value $\alpha_{2}<0$, and the frequency of $Q_{2}$ satisfies $\tilde{\omega}_{2}<\tilde{\omega}_{j}$, thus a switchable NNN coupling can also be realized
with $\alpha_{2}=-\Delta_{1}=-\Delta_{3}$ (see Appendix C for details).

According to Eq.$\,$(3), Figure $2$(a) shows the calculated $J_{1(\textbf{1})}$
and $J_{1(\textbf{0})}$ versus the $Q_{2}$ anharmonicity $\alpha_{2}$ for
system parameters given as: $g_{j}/2\pi=45\,\rm MHz$, $\Delta_{j}/2\pi=-500\,\rm MHz$,
and $\alpha_{j}/2\pi=-350\,\rm MHz$. One can find that when $\alpha_{2}/2\pi$ takes
a value of $500\,\rm MHz$, i.e., $\alpha_{2}=-\Delta_{1(3)}$, the NNN coupling strength
$J_{1(\textbf{1})}$ is zero for $Q_{2}$ in $|1\rangle$, while for $Q_{2}$ in $|0\rangle$,
since the coupling strength $J_{1(\textbf{0})}$ is independent of $\alpha_{2}$, the
interaction is intact, thus allowing us to control the NNN coupling with a high on/off
ratio, where the state of $Q_{2}$ functions as an on/off switch. Therefore, at the
working point $\alpha_{2}=-\Delta_{1(3)}$, the effective NNN coupling can be described
by
\begin{eqnarray}
\begin{aligned}
H_{\rm CXY}=J_{1(\textbf{0})}[|0\rangle\langle 0|]_{2}\otimes\big([|01\rangle\langle 10|]_{1,3}+[|10\rangle\langle 01|]_{1,3}\big).
\end{aligned}
\end{eqnarray}
Similar result can also be obtained for the terms ZZZ and ZIZ in Eq.$\,$(2),
which describe the dispersive ZZ coupling between $Q_{1}$ and $Q_{3}$ resulting
from the virtual exchange interaction between qubit states and non-qubit states,
the interaction strength of which also depends on the state of $Q_{2}$,
thus enabling the ZZ interaction controlled by the state of $Q_{2}$ (see Appendix
D for details).

\section{Realization of Controlled-$\rm i$SWAP gate}

\begin{table}[htbp]
	\centering
	\caption{System parameters used for implementing the C-iSWAP gate.}
\begin{tabular}{cccc}
  \hline\hline
   Qubits & $Q_{1}$ & $Q_{2}$ & $Q_{3}$ \\
   \hline
  Anharmonicity $\alpha/2\pi\,(\rm MHz)$& -350 & 350 & -350 \\

  Idle frequency $\omega_{i}/2\pi\,(\rm GHz)$ & 5.15 & 6.35 & 5.30 \\

  Interaction frequency $\omega_{I}/2\pi\,(\rm GHz)$ & $\sim$6.00 & 6.35 & $\sim$6.00 \\

  NN coupling strength $g/2\pi\,(\rm MHz)$ &  \multicolumn{3}{c}{45  $\,\,\,\,\,$   45}\\

  \hline\hline
\end{tabular}
\end{table}

\begin{figure}[tbp]
\begin{center}
\includegraphics[width=8cm,height=6cm]{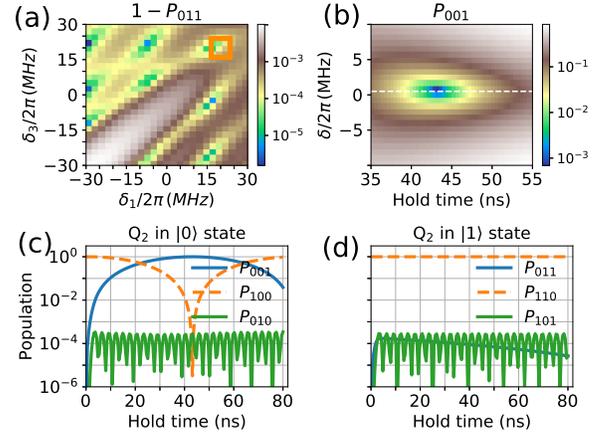}
\end{center}
\caption{Optimal working point for C-iSWAP gate. (a) Swap error $1-P_{011}$ versus
the frequency offset $\delta_{1}$ and $\delta_{3}$ that are defined with respect
to the ideal interaction point, as shown in the inset of Fig.$\,$2(c). The system is
initialized in the eigenstate $|\widetilde{011}\rangle$ at the idle point, and the hold
time takes a value of 45 $\rm ns$. The (orange) square indicates the working point where the
NNN exchange interaction is turned off for $Q_{2}$ in $|1\rangle$. (b) Population $P_{001}$
versus the overshoot $\delta_{s}=\delta_{1}-\delta_{3}$ and hold time for system initialized in the eigenstate
$|\widetilde{001}\rangle$ at the idle point. The horizontal cut (dashed line) depicts
the optimal value of overshot for enabling a full complete swap betweeen
$|\widetilde{001}\rangle$ and $|\widetilde{100}\rangle$. (c) Population swap
$|\widetilde{001}\rangle\Leftrightarrow|\widetilde{100}\rangle$ and (d)
$|\widetilde{011}\rangle\Leftrightarrow|\widetilde{110}\rangle$ versus hold time
for system initialized in $|\widetilde{100}\rangle$ and $|\widetilde{110}\rangle$,
respectively. With optimal frequency offset and overshot obtained from (a) and (b),
the NNN exchange interaction is turned on or off depending on the state of $Q_{2}$.}
\end{figure}

\begin{figure}[tbp]
\begin{center}
\includegraphics[width=8cm,height=8cm]{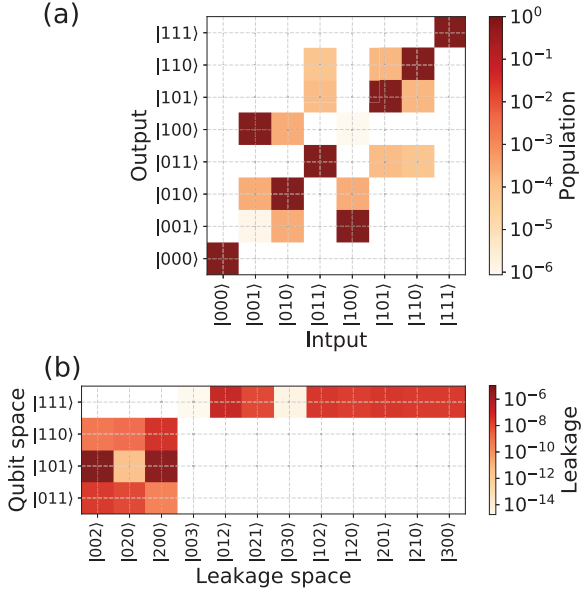}
\end{center}
\caption{ C-iSWAP gate implementation. (a) C-iSWAP gate truth table from
the numerical simulation with the system initialized in the logical basis state.
For each input state taken from logical basis, the output state shows an agreement
with the expected result from the ideal C-iSWAP. (b) Leakage during the
gate implementation from the numerical simulation with the system initialized in
the qubit subspace that is susceptible to leakage error. Leakage to non-qubit state
is suppressed below $10^{-5}$.}
\end{figure}
Having shown the switchable two-qubit exchange coupling, we now turn to use it to demonstrate
controlled two-qubit operations. From Eq.$\,$(4), it becomes clear that the switchable NNN
coupling at the working point could be used to realize the C-iSWAP gate with an arbitrary
swap angle $\theta$, i.e.,
\begin{equation}
U_{CXY}(\theta)=\left(
\begin{array}{cccccccc}
1 & 0 & 0& 0 &0 & 0& 0 &0\\
0 & \cos{\frac{\theta}{2}} & 0 & 0 & i\sin{\frac{\theta}{2}} & 0& 0 &0\\
0 & 0 & 1 & 0& 0 & 0& 0 &0\\
0 & 0 & 0 & 1 & 0 & 0& 0 &0\\
0 & i\sin{\frac{\theta}{2}} & 0 & 0& \cos{\frac{\theta}{2}} & 0& 0 &0\\
0 & 0 & 0 & 0& 0 & 1& 0 &0\\
0 & 0 & 0 & 0& 0 & 0& 1 &0\\
0 & 0 & 0 & 0& 0 & 0& 0 &1\\
\end{array}
\right)
\end{equation}
which becomes a C-iSWAP gate for $\theta=\pi$. However, we note that since this
effective coupling results from dispersive NN coupling (second-order process),
its coupling strength has a magnitude similar to the residual dispersive coupling
between adjacent qubits that are presented in the full system effective Hamiltonian
$H_{\rm eff}$, i.e., ZZ coupling terms $\rm ZZI$ and $\rm IZZ$ \cite{R11} (while terms $\rm ZIZ$
and $\rm ZZZ$ are originated from fourth-order process \cite{R12}, see also in Appendixes B and D).
Hence, these residual interactions impose a limiting factor on the fidelity of the
native C-iSWAP gate.

As demonstrated in the previous study \cite{R14}, in order to suppress the unwanted ZZ coupling
between adjacent qubits in the present system where qubits are coupled together directly
via a capacitor, the anharmonicity of $Q_{2}$ should have a similar magnitude to those of the
two adjacent qubits $Q_{1}$ and $Q_{3}$. This means that the optimal working point for
realizing C-iSWAP gate using the NNN coupling is $\alpha_{2}=-\Delta_{1(3)}=-\alpha_{1(3)}$,
as shown in Fig.$\,$2(b), where the calculated $J_{1(\textbf{1})}$ versus the anharmonicity
$\alpha_{2}$ and the qubit frequency detuning $\Delta_{j}$ with system parameters
$g_{j}/2\pi=45\,\rm MHz$ and $\alpha_{j}/2\pi=-350\,\rm MHz$ is presented according
to Eq.$\,$(3).

Based on the above analysis, in the following discussion, we show a case study that explores
the switchable NNN coupling in a three-qubit system with always-on interaction to
implement the C-iSWAP gate. The system parameters are tabulated in Table $\rm I$. During
the implementation of the three-qubit gate, the frequency of the intermediary qubit $Q_{2}$
is fixed, and the frequencies of the two
NN qubits $Q_{1}$ and $Q_{3}$ vary from the idle frequency point $\omega_{i}$ to the
interaction frequency point $\omega_{I}$ according to a time-dependent function as shown
in Fig.$\,$2(c), where the hold time is defined as the time-interval between the
midpoints of the ramps \cite{R27} (see Appendix E). According to the analytical
derivation of the effective NNN coupling, Figure $\,$2(d) shows the analytical strength of
the exchange interactions $J_{1(\textbf{1})}$ and $J_{1(\textbf{0})}$ as a function of time during
the typical gate implementation with respect to the control pulse shown in Fig.$\,$2(c). We note
that this analytical expression of the exchange interactions is derived based on the basis that
is dressed by the NN coupling between $Q_{2}$ and $Q_{j}$ (the full system Hamiltonian $H$ is written with a
block-diagonal form as $H_{\rm eff}$ in terms of this dressed basis), rather than the logical basis that is
defined as the eigenstate of the full system Hamiltonian $H$ at the parking (idling)
point (see Appendix A).

Before going further, it is important to note that the above analysis, as well as the
optimal working point, is derived based on the analytical expression for
$J_{1(\textbf{1})}$ given in Eq.$\,$(3), which is valid under the dispersive
condition, i.e., $|\Delta_{j}/g_{j}|\gg1$. However, for a system with parameters tabulated
in Table $\rm I$, one may argue that the system operates in a quasi-dispersive
regime at the interaction point, where $|\Delta_{j}/g_{j}|=|\alpha_{2}/g_{j}|\approx7.78$.
Nonetheless, as shown in the following numerical analysis, although operating in the quasi-dispersive
regime, the above results do approximate the full system dynamics well.

In order to find the optimal working point numerically, firstly, according to the expression
given in Eq.$\,$(3), we estimate the time $T$ for realizing a full swap when the system is
initialized in state $|001\rangle$, giving $T\approx\pi/2J_{1(\textbf{0})}= 45\,\rm ns$. Therefore, as shown in
Fig.$\,$3(a), by initializing the system in eigenstate state $|\widetilde{011}\rangle$
at the idling point (Note that at the idling point, the
inter-qubit coupling is effectively turned off, and the logical basis
state is defined as the eigenstates of the system biased
at this point, i.e, $|\widetilde{ijk}\rangle$, which is adiabatically
connected to the bare state $|ijk\rangle$),
and varying the qubit frequencies according to the pulse
shown in Fig.$\,$2(c) with hold time of 45 $\rm ns$, we numerically study the
swap error defined as $1-P_{011}$ ($P_{011}$ denotes the population in $|\widetilde{011}\rangle$
after the time evolution) versus $\delta_{1}$ and $\delta_{3}$ that
are defined as the frequency offsets with respect to the ideal interaction point,
as shown in the inset of Fig.$\,$2(c). In Fig.$\,$3(a), the square indicates the
working point where the two offsets are equal, i.e., $\delta_{1}=\delta_{3}$, thus
preserving on-resonance condition for iSWAP gate, and meanwhile, the swap error is much
smaller than other points on the diagonal of the parameter space, which means that at this point,
the NNN coupling is turned off for $Q_{2}$ in $|1\rangle$ state. To enable a complete
swap in our fixed coupled system that is initialized in state $|\widetilde{001}\rangle$, we
further consider a small frequency overshoot $\delta_{s}=\delta_{1}-\delta_{3}$ applied on $Q_{1}$ \cite{R28},
as shown in the inset of Fig.$\,$2(c), and the horizontal dashed line in Fig.$\,$3(b) depicts the optimal
value of the overshot for this purpose.

Hence, with the optimal frequency offset and overshoot obtained from the above numerical
analysis, Figures $\,$3(c) and 3(d) show the state population versus hold time for system
initialized in eigenstate state $|\widetilde{100}\rangle$ and $|\widetilde{110}\rangle$ at
the idle point, respectively. One can find that for $Q_{2}$ prepared in $|0\rangle$, the
NNN exchange interaction is turned on, enabling an almost complete population swap between $Q_{1}$
and $Q_{3}$ ($|\widetilde{001}\rangle\Leftrightarrow|\widetilde{100}\rangle$), while
for $Q_{2}$ prepared in $|1\rangle$, the exchange interaction is turned off, thus there
is no population swap between $Q_{1}$ and $Q_{3}$. Moreover, although operating in the
quasi-dispersive regime, during the time evolution, the population in $Q_{1}$ or $Q_{3}$
leaking to $Q_{2}$ can still be strongly suppressed, as shown in Figs.$\,$3(c) and 3(d).

As already mentioned before, a direct application of the switchable NNN coupling
demonstrated in Fig.$\,$3 is the implementation of the $U_{\rm CXY}(\theta)$ gate
given in Eq.$\,$(5). Here, for illustration purpose, we consider the implementation of
C-iSWAP gate, i.e., $U_{\rm CXY}(\pi)$, which is realized with a hold time of $43.2\, \rm ns$,
as shown in Fig.$\,$3. By preparing system in eight logical basis states (eigenstates at the
idle point), Figure 4$\,$(a) shows the output basis state, exhibiting good agreement with
the expected result from the ideal C-iSWAP.

\subsection{Intrinsic gate performance}

To quantify the intrinsic gate performance of the implemented C-iSWAP gate,
we consider the average gate fidelity defined as \cite{R29}
\begin{eqnarray}
\begin{aligned}
F=\frac{{\rm Tr}(UU^{\dag})+|{\rm Tr}(U_{CXY}(\pi/2)U^{\dag})|^{2}}{72},
\end{aligned}
\end{eqnarray}
where $U$ is the actual evolution operator (excluding the effect of the decoherence process)
in terms of the logical basis. This is calculated with the full system Hamiltonian of Eq.$\,$(1),
where each qubit are modeled as a four-level system (see Appendix E for more details). Truncated to the qubit
levels, and up to single-qubit phase gates and a global phase \cite{R27,R30,R31} (see also
Appendix E for details), we find that our gate has an intrinsic
fidelity of $F=99.97\%$ for gate time in $50$ $\rm ns$. Aside from the control error,
this high intrinsic gate fidelity is enabled by (i) the
low leakage error, as shown in Fig.$\,$4(b), where one can find that the leakage to non-qubit
state is suppressed below $10^{-5}$, (ii) lower coherence phase error, which is caused by parasitic
ZZ coupling between qubits, as tabulated in Table $\rm II$, the accumulated phase resulting from the
interaction between qubit state and non-qubit is suppressed below $0.02\,\rm rad$.

Furthermore, from Table $\rm II$, one can find that the coherence phases accumulated in
state $|\widetilde{011}\rangle$ and $|\widetilde{110}\rangle$ are smaller than $0.005\,\rm rad$. Considering the
rather strong NN coupling, these rather low accumulated phases demonstrate that by coupling
two-type of qubits with opposite-sign anharmonicities together, the residual ZZ interaction is
heavily suppressed, as shown in previous studies \cite{R13,R14}. And since the coupling between NNN
qubits is enabled by second-order process, and the qubits have a considerably larger anharmonicity,
the ZZ coupling between NNN qubits is also suppressed (note that as mentioned above, the virtual exchange
interactions between qubit state and non-qubit state are dependent on the
state of $Q_{2}$ (see Appendix E), thus the accumulated phases in $|\widetilde{101}\rangle$ and $|\widetilde{111}\rangle$
are different). However, we note that aside from the leakage error, as in the case of
two-qubit iSWAP gate \cite{R28,R32,R33}, this residual ZZ coupling between NNN qubits imposes
a fundamental tradeoff between the fidelity of C-iSWAP gate and the gate speed.

Although the above demonstration has shown that the native implementation of C-iSWAP gate
with high intrinsic fidelity and shorter gate time should be possible with realistic
parameters, we note that these successes are based on a rather strong NN coupling
(although feasible with present technology, but it is larger than the typical NN coupling
commonly used in practice \cite{R4,R34}) and the qubit frequency with a larger tunable range. In practice,
the strong always-on coupling in the present work may make single qubit addressing \cite{R35}, as well
as the implementation of the two-qubit gates \cite{R4,R34}, a challenge for system with limited frequency
tunability. However, we find that with a smaller NN coupling with strength of $30\,\rm MHz$ \cite{R4},
the intrinsic gate fidelity above $99\%$ ($99.9\%$) can still be achieved in 50 (100) $\rm ns$ (see Appendix E).
Moreover, the present protocol could also be applied to the system with tunable NN
coupling \cite{R5,R36,R37,R38}, thus removing the above mentioned constrains.

\begin{table}[htbp]
	\centering
	\caption{Accumulated coherence phase caused by the parasitic ZZ coupling between
      qubits during the gate implementation.}
\begin{tabular}{ccccc}
  \hline\hline
   & 011 $\,\,\,\,\,\,\,\,$ & 101 $\,\,\,\,\,\,\,\,$& 110 $\,\,\,\,\,\,\,\,$& 111\\\hline
  Phase ($\phi\times10^{-3}\,\rm rad$) & -4.40 $\,\,\,\,\,\,\,\,$& -17.28 $\,\,\,\,\,\,\,\,$& -3.67 $\,\,\,\,\,\,\,\,$& 8.67 \\
  \hline\hline
\end{tabular}
\end{table}

\subsection{Impact of decoherence process}

\begin{figure}[tbp]
\begin{center}
\includegraphics[width=8cm,height=6cm]{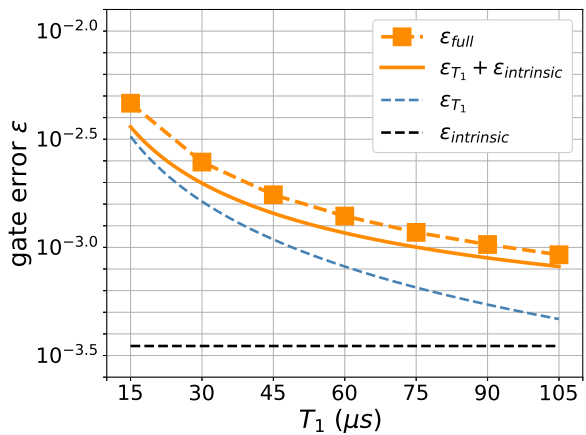}
\end{center}
\caption{Relation between gate infidelity $\varepsilon$ and qubit relaxation time
$T_{1}$ (here assuming that qubit dephasing time $T_{\phi}=\infty$).
Since $T_{1}$ of the C-shun flux qubit can have a magnitude comparable
to that of the Transmon qubit, here $T_{1}$ of the three qubits takes the same value.
$\varepsilon_{\rm full}=1-F_{o}$ denotes the gate infidelity, that is obtained by solving the Lindblad
master equation in the Liouville representation, $\varepsilon=1-F$ represents the intrinsic
gate infidelity, and $\varepsilon_{T_{1}}$ corresponds to the infidelity assuming only $T_{1}$ decay. }
\end{figure}

Here we evaluate the impact of relaxation effect on our proposed gate implementation.
Since the fidelity $F$ defined in Eq.$\,$(6) is no longer valid in the presence
of decoherence process, we will instead use the average gate fidelity $F_{o}$
defined as \cite{R29,R39,R40,R41}
\begin{eqnarray}
\begin{aligned}
F_{o}=\frac{8(1-L_{1})+{\rm Tr}(\mathcal{P}_{U}^{\dag}\mathcal{P}_{U_{\rm{target}}})}{72},
\end{aligned}
\end{eqnarray}
where $\mathcal{P}_{U}$ denotes the system time-evolution superoperator, that is obtained by
solving the Lindblad master equation in the Liouville representation \cite{R42,R43} (see Appendix E for
more details), $\mathcal{P}_{U_{\rm{target}}}=U_{\rm{target}}^{\ast}\otimes U_{\rm{target}}$
(Here $U_{\rm target}=U_{\rm CXY}(\pi/2)$) denotes
the target gate operation in the Liouville
representation, and $L_{1}$ denotes the leakage of the gate
operation, given as \cite{R39,R40}
\begin{eqnarray}
\begin{aligned}
L_{1}=1-\frac{1}{8}\sum_{i,j,k\in\{0,1\}}{\rm Tr}(\mathcal{P}_{U}|ijk\rangle\rangle),
\end{aligned}
\end{eqnarray}
where $|ijk\rangle\rangle$ denotes the logical qubit state $|ijk\rangle$ represented in the Liouville
space. We note that by ignoring the decoherence process, the above defined average gate fidelity $F_{o}$
is consistent with the fidelity $F$ of Eq.$\,$(6) \cite{R29,R39,R40,R41}.

According to the above defined fidelity $F_{o}$, Figure 5 shows the relation between gate
infidelity $\varepsilon_{\rm full}=1-F_{o}$ and qubit relaxation time $T_{1}$ (here assuming that
qubit dephasing time $T_{\phi}=\infty$). We note that at current state, the typical value of $T_{1}$
of the C-shun flux qubit can have a magnitude comparable
to that of the Transmon qubit \cite{R24}, thus $T_{1}$ of the three qubits takes the same value in our
numerical analysis. In Fig.$\,$5, we have also shown the intrinsic gate infidelity $\varepsilon=1-F$ and
the infidelity assuming only $T_{1}$ decay (see Appendix E for more details) \cite{R29}
\begin{eqnarray}
\begin{aligned}
\varepsilon_{T_{1}}\approx3\left[1-\frac{3+e^{-t/T_{1}}+2e^{-t/2T_{1}}}{6}\right].
\end{aligned}
\end{eqnarray}
From Fig.$\,$5, one can find that gate fidelity of $99.5\%$ ($99.9\%$) could be achieved with
qubit relaxation time $T_{1}\geq15\,\mu s$ ($T_{1}\geq105\,\mu s$). In Fig.$\,$5, we also show
that combining the intrinsic gate infidelity and the infidelity assuming only $T_{1}$ decay gives
an estimated value of the gate infidelity $\varepsilon+\varepsilon_{T_{1}}$, which is in
good agreement with $\varepsilon_{\rm full}$.

In the discussion above, we have omitted the effect of the qubit
dephasing process ($T_{\phi}=\infty$). Here, we give an estimate of the effect of the qubit
dephasing process (white noise) on the gate performance. With current
superconducting qubits, the qubit dephasing time (white noise) can commonly reach
up to $T_{\phi}=2T_{1}$ \cite{R44}, giving the infidelity assuming only $T_{1}$ and $T_{\phi}$
process as
\begin{equation}
\begin{aligned}
\varepsilon_{T_{1}}&\approx3\left[1-\frac{3+e^{-t/T_{1}}+2e^{-t(1/2T_{1}+1/T_{\phi})}}{6}\right]
\\&\approx \frac{t}{T_{1}}+\frac{t}{T_{\phi}}.
\end{aligned}
\end{equation}
Thus, one may reasonably estimate that C-iSWAP gate with fidelity
in excess of $99\%$ could be attainable with current technology, and gate infidelity
primarily results from qubit relaxation.

\section{Possible application of the $U_{\rm CXY}$ gates }

Here we discuss two possible applications of our proposed native three-qubit
gates $U_{\rm CXY}(\theta)$, and show that compared with a quantum processor having
only native single- and two-qubit gates, implementing native three-qubit gates
can potentially reduce circuit depth or gate count of quantum
circuits, such as a three-qubit Toffoli gate and controlled-XX (controlled-YY/ZZ)
evolution with arbitrary rotation angle $\theta$ for quantum chemistry
using the quantum phase estimation algorithm (PEA) \cite{R20}. We note that our proposed
controlled two-qubit operation is based on the effective switchable NNN exchange
coupling, which is turned on (off) when the intermediary qubit is at its
ground state (excited state). Thus, following the convention taken in quantum
computing community, one may relabeled the ground state (excited state) of the intermediary
qubit as its logical state $|1\rangle$ ($|0\rangle$).

\subsection{The three-qubit Toffoli gate}

\begin{figure}[tbp]
\begin{center}
\includegraphics[width=8cm,height=6cm]{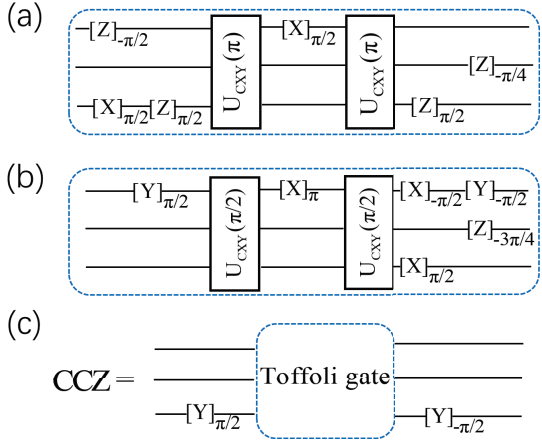}
\end{center}
\caption{Quantum circuit diagram decomposition of the three-qubit Toffoli
gate and CCZ gate with $U_{\rm CXY}(\theta)$ and single qubit
gates $[\mathcal{O}]_{\theta}\equiv\rm{exp}[-i\theta(\mathcal{O}/2)]$.
(a) Implementation of the Toffoli gate with $U_{\rm CXY}(\pi)$ (C-iSWAP gate). (b) Implementation
of the Toffoli gate with $U_{\rm CXY}(\pi/2)$ (C-$\sqrt{\rm{iSWAP}}$ gate). (c) Implementation of CCZ gate with
quantum circuit in (a) or (b).}
\end{figure}

The textbook three-qubit gate Toffoli gate, which is widely
used in various quantum circuits \cite{R19}, in general cannot be implemented
natively \cite{R45,R46,R47} without resorting to the decomposition approach
that involves a series of gate sequences including single- or
two-qubit gates. Even with the decomposition approach, the circuit
depths and gate count needed for implementing Toffoli gate depend
heavily on the available native gate set \cite{R45,R46,R47,R48}. For a fully connected quantum
processor with native gate set including one- and two-qubit gates,
the most common approach for realizing Toffoli gate requires six CNOT
gates or CZ gates and multiple one-qubit gates \cite{R45}, while in a quantum
processor with NN coupling, which is one of the most native architectures for realizing a
scalable superconducting quantum processor, since the available native
two-qubit gate is only possible for pair of NN qubits, more two-qubit gates are
needed on account of this limited connectivity \cite{R45,R48}. However, as
shown in Figs.$\,$6(a) and 6(b), a Toffoli gate can be implemented
via our proposed native controlled two-qubit peration with only two
applications of $U_{\rm CXY}(\pi)$ (a simple extension of the results in Ref.$\,$\cite{R45})
or $U_{\rm CXY}(\pi/2)$ (a simple extension of the results in Ref.$\,$\cite{R49}), thus heavily reducing the
circuit depth and gate count. As shown Figs.$\,$6(c), with two
additional one-qubit gates, the CCZ gate can also be constructed
by using only two $U_{\rm CXY}(\pi)$ or $U_{\rm CXY}(\pi/2)$. Therefore, one
may reasonably estimate that using the above decomposition method with
native three-qubit gates $U_{\rm CXY}(\theta)$ could improve the performance of
the implemented quantum circuits and increase in success probability,
especially, in the NISQ era.

\subsection{Controlled-XX rotation in PEA for quantum chemistry}
\begin{figure}[tbp]
\begin{center}
\includegraphics[width=8cm,height=6cm]{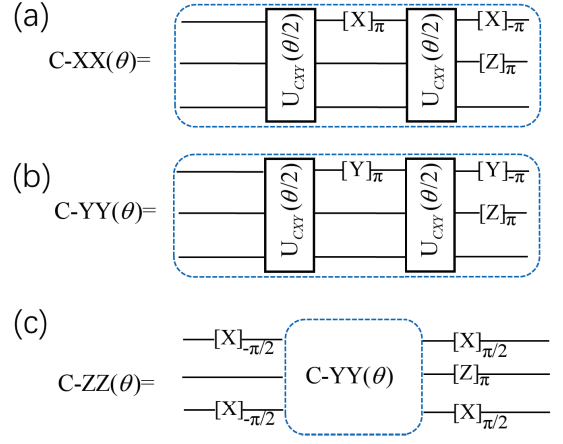}
\end{center}
\caption{Quantum circuit diagram decomposition of the controlled-XX (YY, ZZ) rotation
with an arbitrary angle $\theta$ in terms of $U_{\rm CXY}(\theta/2)$ gates and
single-qubit gates $[\mathcal{O}]_{\theta}\equiv\rm{exp}[-i\theta(\mathcal{O}/2)]$.
(a),(b) Construction of CXX($\theta$) and  CYY($\theta$) with $U_{\rm CXY}(\theta/2)$.
(c) Construction of CZZ($\theta$) with quantum circuit in (b).}
\end{figure}

Figures 7(a) and 7(b) show that based on our proposed three-qubit operation $U_{\rm CXY}(\theta)$,
controlled-XX (controlled-YY) operations with arbitrary rotation angle $\theta$, i.e., ${\rm CXX}(\theta)$,
${\rm CYY}(\theta)$, can be constructed via the application of two $U_{\rm CXY}(\theta/2)$. Moreover, with five
additional one-qubit gates, controlled-ZZ operation can also be implemented as shown in Fig.$\,$7(c).
Having access to these continuous sets of controlled two-qubit rotations may provide more
efficient compiled circuits for quantum application such as quantum simulation
of Fermionic Hamiltonian \cite{R17,R50} and quantum chemistry \cite{R20}. As a simple example we consider
the application of these continuous sets of controlled two-qubit rotations in PEA
for quantum chemistry, where the controlled two-qubit rotations are needed for implementing
trotterized time-evolution operator \cite{R20}. By using the standard decomposition procedure with
only one- and two-qubit gates, a ${\rm CXX}(\theta)$ or ${\rm CYY}(\theta)$
operation requires two CNOT gates and one C-Phase gate with phase angle $\theta$ and six
single qubits to be implemented on quantum processor with NN coupling, yielding a quantum
circuit with gate depth of 5 and gate count of 8 \cite{R20}. Moreover, for a quantum processor with only one type of native
two-qubit gate, e.g., CZ gate, more additional single qubit gates are needed (implementing a CNOT
gate requires one CZ gate and two single qubit gates), thus increasing the circuit depth to
7. As shown in Fig.$\,$7(a), the ${\rm CXX}(\theta)$ can be implemented with gate depth of 4 and
gate count of 4. Thus, the continuous sets of controlled two-qubit rotations $U_{\rm CXY}(\theta)$
presented in this work could be useful for quantum simulation and may provide a more efficient
compilation for certain quantum circuits.

\section{Discussion}

\begin{figure}[tbp]
\begin{center}
\includegraphics[width=8cm,height=8cm]{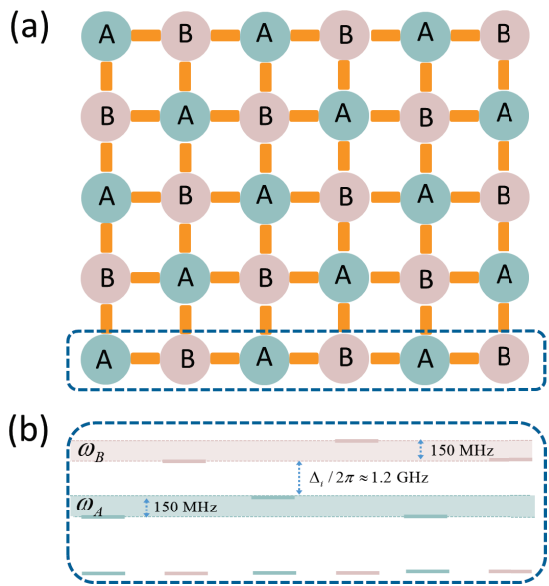}
\end{center}
\caption{(a) A scalable two-dimentional qubit lattice, where circles at the vertices denote
qubits, and yellow lines indicate couplers between adjacent qubits. The lattice
comprises two types of qubits (labeled as A and B) that are arranged in
an -A-B-A-B- pattern in each row and column. (b) Frequency arrangement in
each row and column follows a zigzag pattern with A-type qubit at lower
frequency band ($\omega_{A}$) and B-type qubit at higher frequency
band ($\omega_{B}$).}
\end{figure}

As we have discussed in Sec.$\,$II (see also in Appendix C), the proposed scheme for realizing
switchable NNN coupling can be applied to the -A-B-A- and -B-A-B-type system, where the A-type (B-type)
has a negative (positive) anharmonicity, and the qubit frequency of B-type qubit is larger than that
of the A-type qubit. We also show that one of the most promising implementation of our proposed
superconducting qubit architecture is a superconducting quantum processor comprising transmon qubit (A-type)
and C-shunt flux qubit (B-type) arranged in an -A-B-A-B- pattern. Therefore, the proposed
scheme can also be applied to a two-dimensional (2D) qubit lattice with NN coupling.
An example for 2D qubit lattice with fixed NN coupling is shown in Fig.$\,$8, where
the two-type qubits are arranged in an -A-B-A-B- pattern in each row and column, and
the arrangement of qubit frequency in each row and column follows a zigzag pattern, i.e,
A-type qubit at lower frequency band with typical value of $\omega_{A}$ and B-type qubit at
higher frequency band with typical value of $\omega_{B}$ ($\omega_{B}>\omega_{A}$). As
demonstrated in Sec.$\,$II, at the interaction point (turning on the effective NNN coupling),
frequency detuning $\Delta_{i}=\omega_{B}-\omega_{A}$ approaches the magnitude of the
qubit anharmonicity $|\alpha|$, and the frequencies of the two target NNN qubit is almost
equal to each other. At the parking (idling) point, i.e., turning off the
effective NNN coupling, the frequency detuning $\Delta_{i}$ should be far larger
that the NN coupling strength (typically, $1.2\,\rm GHz$ vs $45\,\rm MHz$) \cite{R4}, as shown in Fig.$\,$8(b).
Moreover, in order to further suppress the residual coupling between NNN qubit (typically
the strength is about $1.7\,\rm MHz$ here), frequency degeneracy in higher or lower frequency
band breaks, yielding a qubit frequency detuning between NNN qubits about $150\,\rm MHz$ .

For practically implementing our scheme in the proposed 2D qubit lattice with fixed NN coupling
shown in Fig.$\,$8, where $A$ (B) label transmon qubit (C-shunt flux qubit), special attention
should be paid to the following two points, (i) the anharmonicity of the transmon qubit is only weakly
dependent on its qubit frequency, thus it is nearly fixed over a tunable range of
approximately $1.0\,\rm GHz$ \cite{R28}. while for the C-shunt flux qubit, the qubit anharmonicity is
strongly dependent on its qubit frequency \cite{R13,R24}. Moreover, in order to suppress parasitic
ZZ interaction between NN qubits, which can lead to conditional phase error during
gate operations \cite{R14}, the C-shunt flux qubit should has an anharmonicity with magnitude
comparable to that of the transmon qubit. This imposes a limitation on the tunable range of
the C-shunt flux qubit in our scheme. Thus, to implement our scheme
in the 2D qubit lattice, the maximum tunable range of each C-shunt flux qubit could be set
below $150\,\rm MHz$, which is adequate for suppressing the parasitic ZZ coupling, and
for providing a sufficient tunable range for implementing high on-off ratio NNN
coupling in a -B-A-B-type system or the 2D lattice shown in Fig.$\,$8(a).
(2) High on-off ratio of the switchable NNN coupling is highly desired for practical purpose.
Since the proposed switchable NNN coupling is enabled by dispersive (quasi-dispersive)
interaction between NN qubits, a strong NN coupling is required for realizing fast gates,
thus reducing the decoherence error. Meanwhile, a larger qubit detuning between
NN qubits is needed for turning it off effectively. Overall, the two points imply that a larger
tunable range of the transmon qubit is needed for practical implementation of our scheme.
However, for transmon qubit far away from its maximum frequency point (flux insensitive
point), the coherence time is commonly heavily suppressed (due to the influence of $1/f$
magnetic flux noise), causing gate infidelity. Hence, the speed-fidelity tradeoff needs to be understood
carefully for implementing our scheme in the proposed 2D qubit lattice with fixed NN
coupling. It is worth noting that although a fixed NN coupling is assumed in the demonstration of
the switchable NN coupling in Set.$\,$II and in the above discussion on the scalability
of our scheme, the proposed scheme is
obviously also compatible with the qubit lattice with tunable NN coupling \cite{R5,R36,R38},
thus relaxing above mentioned constraints.

Finally, we note that our proposed scheme for implementing native C-iSWAP gate is based on
effective three-qubit interaction resulting from NN transverse (XY)
coupling, and the control pulse is described by only a few parameters (e.g., frequency offset $\delta_{j}$,
frequency overshoot $\delta$, and hold time, as shown in Fig.$\,$2), making the \emph{in situ} experimental
optimization of pulse parameters as feasible as the case for the typical one- or two-qubit
gate operations, for which the \emph{in situ} optimization of pulse parameters has been
shown as a key step towards high-fidelity gate operations \cite{R51,R52}. For system with NN transverse coupling,
an alternative (single-shot) scheme for implementing C-iSWAP gate without resorting to the
standard decomposition procedure has also been proposed previously by using quantum optimal
control \cite{R17}, where control pulses are discretized into pixels, and in principle can generate
gates with speed approaching the quantum speed limit. However, the success of the optimal control is
based on the assumption that accurate knowledge of the practical system
parameters is available for optimizing control pulse in theoretical simulation,
and for superconducting qubits, this is often not the case in reality \cite{R51,R52,R53}.
Moreover, the \emph{in situ} optimization of pulse parameters is hindered
by the larger number of pixels, which in principle could be overcome by using
simple parameterizations for the control pulse, such as Fourier and erf
parameterizations \cite{R53}. Recently, another scheme for implementing native C-iSWAP gate
was proposed for superconducting qubit systems with dedicated designed NN coupling \cite{R18}, i.e.,
combining the NN transverse (XY) coupling and the NN longitude (ZZ) coupling, which seems a highly
designed asymmetry architecture, potentially limiting the compatibility with
existing scheme for realizing high-fidelity one- or two-qubit operations. Therefore, this scheme
may be more suitable for certain types of quantum computing applications \cite{R18}.

\section{Conclusion}

In this work, we employ a scalable chain of nearest-neighbor-coupled
superconducting qubit system comprising two-type superconducting qubits with
opposite-sign anharmonicity to realize switchable coupling between
next-nearest-neighboring (NNN) qubits. This switchable coupling is controlled by the
state of the intermediary qubit, thus potentially enabling the implementation
of various native controlled two-qubit operations. With realistic parameters,
we show that it is possible to realize a C-iSWAP gate with an intrinsic average
fidelity of $99.9\,\%$ in 50 $\rm ns$. These native implementations of
three-qubit gates may find useful application for reducing circuit depth
of NISQ algorithms \cite{R6}, such as quantum chemistry \cite{R20}, and for performing
quantum simulations \cite{R49,R54}.

\acknowledgements
We would like to thank Yu Song for helpful suggestions on the manuscript.  This work
was partly supported by the National Key Research and Development Program of China (Grant No.
2016YFA0301802), the National Natural Science Foundation of
China (Grant No. 61521001, and No. 11890704), and the Key R$\&$D Program of
Guangdong Province (Grant No.2018B030326001). P. X. acknowledges the supported by Scientific
Research Foundation of Nanjing University of Posts and Telecommunications (NY218097),
NSFC (Grant No. 11847050), and the Young fund of Jiangsu Natural Science
Foundation of China (Grant No. BK20180750). H. Y. acknowledges support from the
Beijing Natural Science Foundation (Grant No.Z190012).

P. Z. and P. X. contributed equally to this work.

\appendix

\section{effective Hamiltonian for qubit system}

As described in the main text, we consider a linear chain of three
superconducting qubits with nearest-neighbor coupling, and here each qubit
is treated as an ideal two-level system, thus described by the Hamiltonian $H=H_{0}+V$ with
$H_{0}=\sum_{l=1}^3\tilde{\omega}_{l}\sigma_{l}^{z}/2$ and
\begin{eqnarray}
\begin{aligned}
V=\sum_{j=1,3}g_{j}(\sigma_{j}^{+}\sigma_{2}^{-}+\sigma_{j}^{-}\sigma_{2}^{+}),
\end{aligned}
\end{eqnarray}
where $\sigma_{j}^{z,\pm}$ denote the Pauli operators associated with the $j$th
qubit labeled as $Q_{j}$ with bare qubit frequency $\tilde{\omega}_{j}$,
and $g_{j}$ represents the coupling strength between
nearest neighbor qubits $Q_{j}$ and $Q_{2}$.

We now turn to derive an effective Hamiltonian for this three-qubit system, and start
from the original Hamiltonian $H$, which is composed of
an unperturbed part $H_{0}$ with known eigenvalues and eigenstates and a small perturbation part
$V$. We consider that the three-qubit system operates in the dispersive regime, where
the detuning between nearest-neighbor qubit pair is larger than the coupling
strengths between them, thus $|\Delta _{j}|=|\tilde{\omega} _{j}-\tilde{\omega}
_{2}|\gg g_{j}$.

For a system operating in the dispersive coupling regime, we can eliminate the
direct qubit-qubit coupling $V$ via a unitary transformation \cite{R25}
\begin{eqnarray}
\begin{aligned}
H_{\textrm{eff}}=\exp(-X)H\exp(X),
\end{aligned}
\end{eqnarray}
where $X$ is chosen such that the direct coupling between the nearest-neighbor qubit
pairs in the transformed Hamiltonian disappears. By choosing
\begin{eqnarray}
\begin{aligned}
X=\frac{g_{1}}{\Delta_{1}}(\sigma_{1}^{-}\sigma_{2}^{+}-\sigma_{1}^{+}\sigma_{2}^{-})
+\frac{g_{3}}{\Delta_{3}}(\sigma_{3}^{-}\sigma_{2}^{+}-\sigma_{3}^{+}\sigma_{2}^{-}),
\end{aligned}
\end{eqnarray}
one can prove that it satisfies $[H_{0},X]=-V$. Expanding to the second
order of the small parameters ($\frac{g_{j}}{\Delta _{j}}=O(\lambda)$) yields
\begin{eqnarray}
\begin{aligned}
H_{\textrm{eff}}&=H_{0}+\frac{1}{2}[V,X]+O(\lambda^{3})
\\&\approx\sum_{l=1}^3\frac{\omega_{l}}{2}\sigma_{l}^{z}
+J(\sigma_{1}^{+}\sigma_{3}^{-}+\sigma_{1}^{-}\sigma_{3}^{+})\sigma_{2}^{z},
\end{aligned}
\end{eqnarray}
with
\begin{eqnarray}
\begin{aligned}
&\omega_{1}=\tilde{\omega}_{1}+\frac{g_{1}^{2}}{\Delta_{1}},
\\&\omega_{2}=\tilde{\omega}_{2}-(\frac{g_{1}^{2}}{\Delta_{1}}+\frac{g_{3}^{2}}{\Delta_{3}}),
\\&\omega_{3}=\tilde{\omega}_{3}+\frac{g_{3}^{2}}{\Delta_{3}},
\\&J=-\frac{g_{1}g_{3}}{2}(\frac{1}{\Delta_{1}}+\frac{1}{\Delta_{3}}),
\end{aligned}
\end{eqnarray}
where $\omega_{l}$ are the dressed qubit frequencies of $Q_{l}$,
and $J$ is the effective three-qubit interaction strength (next-nearest neighbor coupling)
given as $J=-g_{1}g_{3}(\Delta_{1}+\Delta_{3})/(2\Delta_{1}\Delta_{3})$. From Eq.$\,$(A4), one
can find that the magnitude of the dispersive XY interaction between $Q_{1}$ and $Q_{3}$ is
independent of the state of $Q_{2}$, while the sign of the interaction is set by the $Q_{2}$
state, i.e., for $Q_{2}$ in $|0\rangle$ or $|1\rangle$, the value of the interaction strength
has an opposite sign. Thus, the effective interaction between $Q_{1}$ and $Q_{3}$ is
in fact a three-body interaction.

We note that the original Hamiltonian $H$ is written in the usual bare
basis of uncoupled system eigenstates, i.e., eigenstates of the unperturbed
Hamiltonian $H_{0}$, but the derived effective Hamiltonian $H_{\rm eff}$ given
in Eq.$\,$ (A4) is written in the transformed bare basis defined by the unitary
transformation. In fact, for fixed coupled system, quantum information
processing is commonly performed in the transformed basis \cite{R3,R27}, i.e., eigenstates
of the idle system Hamiltonian $H$, where qubits are all dispersively coupled
to each other, i.e., the strength of the direct or indirect coupling
between arbitrary pair of qubits is far smaller than the frequency detuning of the
coupled qubits. In this way, at the idle point and in the interaction picture,
the system states suffer no dynamic evolution. Therefore, throughout this work,
we take all these factors into consideration, implicitly.

\section{effective Hamiltonian for qubit system with higher energy levels}

Since the superconducting qubit is naturally a multi-level system, especially
for qubits with weak anharmonicity such as the transmon qubit and the C-shunted flux qubit,
the higher energy levels of qubits have non-negligible effect on the effective coupling
derived in the above section where superconducting qubits are treated as an ideal two-level
system \cite{R21,R24}. In the following discussion, we will study the effect of the higher energy
levels of qubits on the dispersive next-nearest neighbor coupling.

For a system consisting of three superconducting qubits (labeled as $Q_{1,2,3}$) as described
in the main text, they can be modeled by a chain of three weakly anharmonic oscillators
with nearest neighbor coupling \cite{R21,R24}. Thus, the Hamiltonian of this system can be described
by $H=H_{0}+V$, with
\begin{eqnarray}
\begin{aligned}
&H_{0}=\sum_{l=1}^3\bigg[\tilde{\omega}_{l}q_{l}^{\dagger}q_{l}+\frac{\alpha_{l}}{2}q_{l}^{\dagger}q_{l}(q_{l}^{\dagger}q_{l}-1)\bigg]
\\&V=\bigg[g_{1}(q_{1}^{\dagger}q_{2}+q_{1}q_{2}^{\dagger})+g_{3}(q_{3}^{\dagger}q_{2}+q_{3}q_{2}^{\dagger})\bigg],
\end{aligned}
\end{eqnarray}
where the subscript $l={1,2,3}$ labels superconducting qubit $Q_{l}$ with
anharmonicity $\alpha_{l}$ and bare qubit frequency $\tilde{\omega}_{l}$,
$q_{l}\,(q_{l}^{\dagger})$ is the associated annihilation (creation)
operator truncated to the lowest four levels (labeled as $\{|0\rangle,|1\rangle,|2\rangle,|3\rangle\}$),
and $g_{j}$ ($j=1,3$) denotes the strength of the coupling between adjacent qubits, i.e., $Q_{j}$
and $Q_{2}$. Again, in the following discussion, we consider that the system operates in the
dispersive regime where the qubit frequency detuning ($\Delta_{j}=\tilde{\omega}_{j}-\tilde{\omega}_{2}$)
is far larger than the NN coupling strength, i.e., $|\Delta_{j}|\gg g_{j}$.

\subsection{Schrieffer-Wolff transformation}

To derive an effective Hamiltonian for the three-qubit system, we turn to
block-diagonalize the original system Hamiltonian $H=H_{0}+\lambda V$, where $\lambda$
is introduced to show the orders in the perturbation expansion, and would be set
to $1$ after the calculations, thus the nearest neighbor coupling between
qubits is eliminated and the next-nearest neighbor qubits are directly coupled
to each other. By projecting the system onto the zero-excitation and one-excitation
subspace of $Q_{2}$, we can further derive an effective Hamiltonian for the
next-nearest neighbor qubits with the intermediary qubit in its ground state $|0\rangle$ or
excited state $|1\rangle$. This is achieved by using the Schrieffer-Wolff transformation \cite{R25,R26}
\begin{eqnarray}
\begin{aligned}
&\,\,\,\,\,\,\,H_{\rm eff}=A^{\dagger}HA
\\&A=e^{-iS},\,\,S=\sum_{n=1}^{\infty}S^{(n)}\lambda^{n}
\end{aligned}
\end{eqnarray}
Following the methods introduced in Ref.\cite{R26}, the effective block-diagonal Hamiltonian
for the three-qubit system has the following form
\begin{equation}
H_{\rm eff}=\left(
\begin{array}{ccccc}
H_{0} & \textbf{0} & \textbf{0}& \textbf{0} &\ldots \\
 \textbf{0} & H_{1} & \textbf{0}& \textbf{0} &\ldots\\
  \textbf{0} & \textbf{0}& H_{2} & \textbf{0} &\ldots \\
  \textbf{0} & \textbf{0}& \textbf{0} & H_{3} &\ldots \\
  \vdots & \vdots& \vdots & \vdots  & \ddots \\
\end{array}
\right);
\end{equation}
where $H_{\rm \textbf{n}}$ ($\rm n = 0, 1, 2, 3, ...$) denotes the effective Hamiltonian for
the three-qubit system projected onto the n-excitation subspace of $Q_{2}$.
Consequently, $H_{0}$ corresponds to the effective Hamiltonian projected onto the
zero-excitation subspace of $Q_{2}$, i.e., the effective Hamiltonian for
the next-nearest neighbor qubits with $Q_{2}$ in state $|0\rangle$. Truncated
to the first three energy levels of qubits, i.e. operating with the basis $\{|000\rangle,\,|001\rangle,\,|100\rangle,\,|101\rangle,\,|002\rangle,\,|200\rangle\}$,
$H_{\textbf{0}}$ reads
\begin{equation}
H_{\textbf{0}}=\left(
\begin{array}{cccccc}
0 &  0 & 0 & 0 & 0&0\\
0 & \omega'_{3} & J_{1(\textbf{0})} & 0 & 0 & 0\\
0 &  J_{1(\textbf{0})} & \omega'_{1} & 0 & 0 & 0 \\
0 &  0 & 0 & \omega'_{3}+\omega'_{1} & J_{2(\textbf{0}),I} & J_{2(\textbf{0}),II}\\
0 &  0 & 0 & J_{2(\textbf{0}),I} & 2\omega''_{3}+\alpha_{3} & 0 \\
0 &  0 & 0 & J_{2(\textbf{0}),II} & 0 & 2\omega''_{1}+\alpha_{1}\\
\end{array}
\right)
\end{equation}
where $\omega'_{j}=\tilde{\omega}_{j}+g_{j}^{2}/\Delta_{j}$ ($j=1,3$) denotes the
dressed transition frequency of $Q_{j}$,
and $\omega''_{j}=\tilde{\omega}_{j}+g_{j}^{2}/(\Delta_{j}+\alpha_{j})$
is defined as an effective dressed qubit frequency, to which the coupling involved
with higher energy levels of qubits contributes with an additional
term $g_{j}^{2}/(\Delta_{j}+\alpha_{j})$.
\begin{eqnarray}
\begin{aligned}
J_{1(\textbf{0})}=\frac{g_{1}g_{3}(\Delta_{1}+\Delta_{3})}{2\Delta_{1}\Delta_{3}}
\end{aligned}
\end{eqnarray}
represents the strength of coupling within one-excitation manifold $\{|001\rangle$, $|100\rangle\}$ of
the NNN qubits, and
\begin{equation}
\begin{aligned}
&J_{2(\textbf{0}),I}=\frac{\sqrt{2}g_{1}g_{3}(\Delta_{1}+\Delta_{3}+\alpha_{3})}{2\Delta_{1}(\Delta_{3}+\alpha_{3})},
J_{2(\textbf{0}),II}\equiv J_{2(\textbf{0}),I}(1\leftrightarrow 3),
\end{aligned}
\end{equation}
corresponds to the strength of coupling within the two-excitation manifold of NNN qubits
$\{|101\rangle$, $|200\rangle$, $|002\rangle\}$, i.e,
the interaction $|101\rangle\leftrightarrow |200\rangle$ and the interaction
$|101\rangle\leftrightarrow |002\rangle$.

$H_{\textbf{1}}$ represents the effective Hamiltonian projected onto the one-excitation
subspace of $Q_{2}$, i.e., the effective Hamiltonian for the next-nearest neighbor
qubits with $Q_{2}$ in state $|1\rangle$. Then, truncated to the first three energy levels
of qubits, i.e., operating with basis
$\{|010\rangle,\,|011\rangle,\,|110\rangle,\,|111\rangle,\,|012\rangle,\,|210\rangle\}$, $H_{\textbf{1}}$
is given as
\begin{equation}
H_{\textbf{1}}=\left(
\begin{array}{cccccc}
E_{010} &  0 & 0 & 0 & 0&0\\
0 &  E_{011} & J_{1(\textbf{1})} & 0 & 0 & 0\\
0 &J_{1(\textbf{1})} & E_{110} & 0 & 0 & 0 \\
0 &0 & 0 & E_{111} & J_{2(\textbf{1}),I} & J_{2(\textbf{1}),II}\\
0 &0 & 0 & J_{2(\textbf{1}),I} & E_{012} & 0 \\
0 &0 & 0 & J_{2(\textbf{1}),II} & 0 & E_{210}\\
\end{array}
\right)
\end{equation}
with
\begin{eqnarray}
\begin{aligned}
&E_{010}=\omega'_{2},
\\&E_{011}=\omega'_{2}+\omega'_{3}+\zeta'_{3},
\\&E_{110}= \omega'_{2}+\omega'_{1}+\zeta'_{1},
\\&E_{111}=\omega'_{2}+\omega'_{3}+\omega'_{1}+\zeta'_{3}+\zeta'_{1},
\\&E_{012}=\omega'_{2}+2\omega'_{3}+\alpha_{3}+\delta_{3},
\\&E_{210}=\omega'_{2}+2\omega'_{1}+\alpha_{1}+\delta_{1},
\end{aligned}
\end{eqnarray}
where $\omega'_{2}=\tilde{\omega}_{2}-g_{1}^{2}/\Delta_{1}-g_{3}^{2}/\Delta_{3}$
denotes the dressed transition frequency of $Q_{2}$, $\zeta'_{j}=2g_{j}^{2}(\alpha_{j}+\alpha_{2})/[(\Delta_{j}-\alpha_{2})(\Delta_{j}+\alpha_{j})]$
represents the parasitic $ZZ$ interaction between adjacent qubits $Q_{j}$ ($j=1,3$) and $Q_{2}$, which
results from the interaction among higher energy levels of qubits, and
\begin{eqnarray}
\begin{aligned}
&\delta_{j}=\frac{g_{j}^{2}(5\alpha_{j}+\Delta_{j}+3\alpha_{2})}{(2\alpha_{j}+\Delta_{j})
 (\Delta_{j}+\alpha_{j}-\alpha_{2})}-\frac{g_{j}^{2}}{\Delta_{j}},
\end{aligned}
\end{eqnarray}
represents frequency shift resulting from the interaction among higher energy levels
of qubits. $J_{1(\textbf{1})}$ describes the strength of coupling within the
one-excitation manifold of next-nearest-neighbor qubits $\{|110\rangle$, $|011\rangle\}$,
and is given as
\begin{eqnarray}
\begin{aligned}
J_{1(\textbf{1})}=\frac{g_{1}g_{3}}{2}\left[\frac{\Delta_{1}+\alpha_{2}}{\Delta_{1}
(\Delta_{1}-\alpha_{2})}+\frac{\Delta_{3}+\alpha_{2}}{\Delta_{3}(\Delta_{3}-\alpha_{2})}\right],
\end{aligned}
\end{eqnarray}
while $J_{2(\textbf{1}),I(II)}$ corresponds to the strength of coupling
within the two-excitation manifold of next-nearest-neighbor qubits
$\{|111\rangle$, $|012\rangle$, $|210\rangle\}$, i.e,
the interaction $|111\rangle\leftrightarrow |012\rangle$ and the interaction
$|111\rangle\leftrightarrow |210\rangle$, and it is given as
\begin{widetext}
\begin{equation}
\begin{aligned}
J_{2(\textbf{1}),I}=\frac{\sqrt{2}g_{1}g_{3}\left(2\Delta_{1}(\Delta_{1}^{2}-\alpha_{2}^{2})
+(3\Delta_{1}^{2}-\alpha_{2}^{2})(\alpha_{3}-\Delta_{1}+\Delta_{3})+
(\alpha_{2}+\Delta_{1})(\alpha_{3}-\Delta_{1}+\Delta_{3})^{2}\right)}
{2\Delta_{1}(\Delta_{1}-\alpha_{2})
 (\Delta_{3}+\alpha_{3})(\Delta_{3}-\alpha_{2}+\alpha_{3})}
,J_{2(\textbf{1}),II}\equiv J_{2(\textbf{1}),I}(1\leftrightarrow 3).
\end{aligned}
\end{equation}
\end{widetext}

From the above result, we can find that: (i) when the anharmonicity of $Q_{2}$
takes a value of $\alpha_{2}=0$, i.e., $Q_{2}$ is a linear resonator, and the
expression of $J_{1(\textbf{1})}$ and
$J_{2(\textbf{1}),I(II)}$ given above will be reduced to $J_{1(\textbf{0})}$ and
$J_{2(\textbf{0}),I(II)}$, respectively, agreeing with the well-known fact that the
liner-bus mediated coupling between two qubits is a two-body interaction between
qubits. (ii) while for $\alpha_{2}\rightarrow \infty$, $Q_{2}$ can
be safely treated as an ideal two-level system, thus $J_{1(\textbf{1})}=-J_{1(\textbf{0})}$
is obtained, as the result shown in Appendix $\,$A.

\subsection{Effective three-qubit Hamiltonian}

For $Q_{2}$ in its ground state $|0\rangle$, the effective interaction among two-excitation
manifold of the pair of nearest-neighboring qubits $Q_{1}$ and $Q_{3}$, i.e.,
interaction between $|101\rangle$ and $|002\rangle\,(|200\rangle)$ as shown in Fig.$\,$9(a), the
strength of which is given in Eq.$\,$(B6), causes the $ZZ$ interaction between $Q_{1}$ and $Q_{3}$
with strength given as
\begin{eqnarray}
\begin{aligned}
\zeta_{101}=\frac{J_{2(\textbf{0}),I}^{2}}{\Delta-\alpha_{3}}
-\frac{J_{2(\textbf{0}),II}^{2}}{\Delta+\alpha_{1}},
\end{aligned}
\end{eqnarray}
where $\Delta=\omega_{1}-\omega_{3}$. While for $Q_{2}$ in its excited state $|1\rangle$, the
strength of the ZZ coupling (resulting from the interaction between
$|012\rangle \,(|210\rangle$) and $|111\rangle$, as shown in Fig.$\,$9(b)) is
\begin{equation}
\begin{aligned}
\zeta_{111}=\frac{J_{2(\textbf{1}),I}^{2}}{\Delta-\alpha_{3}+\zeta'_{1}+\zeta'_{3}-\delta_{3}}
-\frac{J_{2(\textbf{1}),II}^{2}}{\Delta+\alpha_{1}-\zeta'_{1}-\zeta'_{3}+\delta_{1}}.
\end{aligned}
\end{equation}
Truncated to the qubit levels, the effective Hamiltonian of the full system has
the following approximate form
\begin{eqnarray}
\begin{aligned}
H_{\rm eff}=&\omega_{1}\frac{ZII}{2}+\omega_{2}\frac{IZI}{2}+\omega_{3}\frac{IIZ}{2}
+\zeta_{1}\frac{ZZI}{2}+\zeta_{3}\frac{IZZ}{2}
\\&+\frac{J_{1(\textbf{1})}-J_{1(\textbf{0})}}{2}\frac{XZX+YZY}{2}
\\&+\frac{J_{1(\textbf{1})}+J_{1(\textbf{0})}}{2}\frac{XIX+YIY}{2}
\\&+\frac{\zeta_{111}-\zeta_{101}}{4}\frac{ZZZ}{2}+\frac{\zeta_{111}+\zeta_{101}}{4}\frac{ZIZ}{2},
\end{aligned}
\end{eqnarray}
where $\rm(X,Y,Z,I)$ represent the Pauli operator and identity operators, and
the order indexes the qubit number, $\omega_{l}$ denotes dressed qubit frequency
given as
\begin{eqnarray}
\begin{aligned}
&\omega_{1}=\omega'_{1}+\frac{\zeta'_{1}}{2}+\frac{\zeta_{111}+\zeta_{101}}{4},
\\&\omega_{2}=\omega'_{2}+\frac{\zeta'_{1}+\zeta'_{3}}{2}+\frac{\zeta_{111}-\zeta_{101}}{4},
\\&\omega_{3}=\omega'_{3}+\frac{\zeta'_{3}}{2}+\frac{\zeta_{111}+\zeta_{101}}{4},
\end{aligned}
\end{eqnarray}
$\zeta_{j}$ represents the strength of ZZ coupling between adjacent qubits, i.e., $Q_{1(3)}$
and $Q_{2}$, which is given as
\begin{eqnarray}
\begin{aligned}
&\zeta_{1}=\frac{\zeta'_{1}}{2}+\frac{\zeta_{111}-\zeta_{101}}{4},
\zeta_{3}=\frac{\zeta'_{3}}{2}+\frac{\zeta_{111}-\zeta_{101}}{4}.
\end{aligned}
\end{eqnarray}
Taking
\begin{eqnarray}
\begin{aligned}
&J_{Z}=\frac{J_{1}^{(\textbf{1})}-J_{1}^{(\textbf{0})}}{2},
J_{I}=\frac{J_{1}^{(\textbf{1})}+J_{1}^{(\textbf{0})}}{2},
\\&\zeta_{Z}=\frac{\zeta_{111}-\zeta_{101}}{4},\zeta_{I}=\frac{\zeta_{111}+\zeta_{101}}{4},
\end{aligned}
\end{eqnarray}
we recover the effective Hamiltonian of Eq.$\,$(2) of the main text.

In Eq.$\,$(B14), the term associated with $XIX+YIY$ causes the excitation to be
swapped between the next-nearest-neighboring qubits (iSWAP), and to which the
term associated with $XZX+YZY$ contributes with a swap rate whose sign depends on the
state of $Q_{2}$. The terms $ZIZ$ and $ZZZ$ correspond
to the ZZ interaction between the next-nearest-neighbor qubits, i.e., $Q_{1}$
and $Q_{3}$, which is resulting from the virtual exchange interaction between qubit
state and non-qubit states, as shown in Fig.$\,$9.

\begin{figure}[tbp]
\begin{center}
\includegraphics[width=8cm,height=2.5cm]{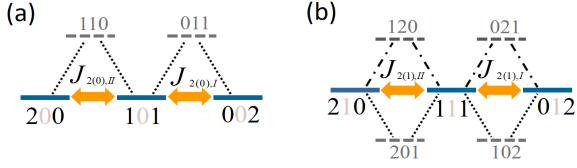}
\end{center}
\caption{Virtual exchange interaction (denoted as the double-headed arrows) among
the two-excitation subspace of the two NNN qubits. (a) For the intermediary qubit
$Q_{2}$ in $|0\rangle$ state, the NNN exchange interaction
$|101\rangle\leftrightarrow|002\rangle$ with strength $J_{2(\textbf{0}),I}$ is enabled by
a single path (denoted as the dashed lines) involving the intermediate state
$|011\rangle$. (b) For $Q_{2}$ in $|1\rangle$ state, the NNN interaction $|111\rangle\leftrightarrow|012\rangle$
with strength $J_{2(\textbf{1}),I}$ is enabled by two paths (denoted as dashed lines and
dash-dotted lines, respectively), and each involves an intermediate state, i.e.,
$|102\rangle$ or $|021\rangle$. Similar results are also obtained for the interaction
$|101\rangle\leftrightarrow|200\rangle$ with strength $J_{2(\textbf{0}),II}$
and the interaction $|111\rangle\leftrightarrow|210\rangle$ with strength $J_{2(\textbf{1}),II}$. }
\end{figure}

\section{Switchable NNN coupling for -B-A-B-type system}

\begin{figure}[tbp]
\begin{center}
\includegraphics[width=8cm,height=4cm]{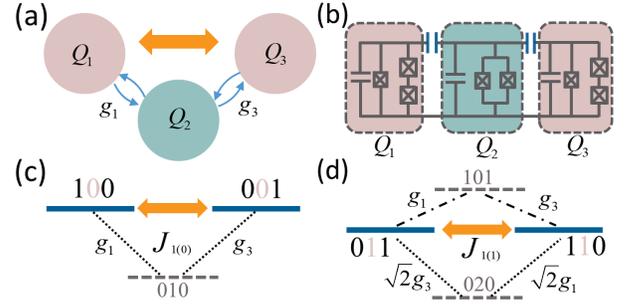}
\end{center}
\caption{(a) Sketch of a -B-A-B-type system with NN coupling. Dispersive interaction
between adjacent qubits (denoted as round arrows) can result in an effective
NNN coupling (denoted as double-headed arrows). (b) Circuit diagram of a chain of
three superconducting qubits capacitively coupled to each other, where the $Q_{1(3)}$
and $Q_{2}$ are C-shunted flux qubit and transmon qubit, which can be modeled as weak
anharmonic oscillators with positive and negative anharmonicity, respectively. (c)
For the intermediary qubit $Q_{2}$ in $|0\rangle$ state, the NNN exchange interaction
with strength $J_{1(\textbf{0})}$ is enabled by a single path (denoted as dashed lines)
involving the intermediate state $|010\rangle$. (d) For $Q_{2}$ in
$|1\rangle$ state, the NNN interaction with strength $J_{1(\textbf{1})}$
is enabled by two paths (denoted as dashed lines and dash-dotted lines, respectively),
and each involves an intermediate state, i.e.,$|101\rangle$ or $|020\rangle$.}
\end{figure}

As discussion in Sec.$\,$II of the main text, and also according to Eqs.$\,$(B5) and (B10) (Eq.$\,$(3)
in the main text), one can find that in a three-qubit system with NN coupling, when $\alpha_{2}=-\Delta_{1}=-\Delta_{3}$,
the two competitive contributions in Eq.$\,$(B10) yield zero net coupling strength
$J_{1(\textbf{1})}=0$, and the $J_{1(\textbf{0})}$ is preserved. Thus, a switchable NNN coupling can be
realized in two concrete settings (Here note that in present work, the A (B) labels qubit with
negative (positive) anharmonicity, and a promising implantation is the
transmon-CSFQ (C-shunt flux qubit)-transmon system): (1) in an -A-B-A-type setting
as shown in Figs.$\,$1(a) and 1(b) , the anharmonicity of $Q_{2}$ (B-type qubit)
takes a positive value. Thus, in order to meet the requirement for implementing the
switchable NNN coupling, i.e., $\alpha_{2}=-\Delta_{1}=-\Delta_{3}$,
the qubit frequency of $Q_{2}$ should satisfy $\tilde{\omega}_{2}>\tilde{\omega}_{j}$. This has been
demonstrated in the main text. (2) in a -B-A-B-type setting (CSFQ-transmon-CSFQ) as shown in Figs.$\,$10(a)
and 10(b), the anharmonicity of $Q_{2}$ (here is an A-type qubit) takes a negative value. As shown in Fig.$\,$10(d),
in order to destructively interfere the two terms in Eq.$\,$(B10), the qubit frequency of $Q_{2}$ should
satisfy $\tilde{\omega}_{2}<\tilde{\omega}_{j}$. Hence, when $\alpha_{2}=-\Delta_{1}=-\Delta_{3}$, a switchable
coupling can also be realized in this setting.

\section{Switchable NNN exchange interaction for higher energy levels}
\begin{figure}[tbp]
\begin{center}
\includegraphics[width=8cm,height=6cm]{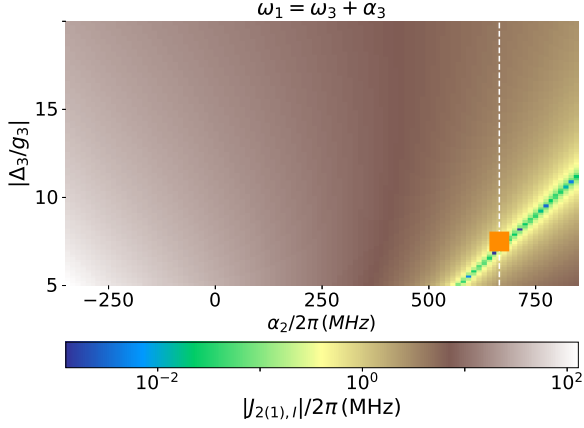}
\end{center}
\caption{Calculated
values of $J_{2(\textbf{1}),I}$ versus $\alpha_{2}$ and $\Delta_{3}$ in the unit of $g_{3}$
with $\omega_{1}=\omega_{3}+\alpha_{3}$. The anharmonicity $\alpha_{j}$ and NN coupling strength
$g_{j}$ take the same values as in Fig.$\,$2(b) of the main text. The intersection (orange
square) of the vertical cut and the dark strip gives
the ideal optimal working point for realizing switchable coupling between $|11\rangle_{1,3}$
and $|02\rangle_{1,3}$ with $\alpha_{2}/2\pi$ take a value of $665\,\rm MHz$.}
\end{figure}

From the expression in Eqs.$\,$(B6) and (B11), one can find that a switchable exchange
interaction between qubit state and non-qubit states for $Q_{1}$ and $Q_{3}$, such as
the interaction $|11\rangle_{1,3}\leftrightarrow|02\rangle_{1,3}$ with strength
$J_{2(\textbf{1}),I}$, can be achieved by
engineering the anharmonicity of $Q_{2}$, as shown in Fig.$\,$9. Hence, as same as the case
for the virtual exchange interaction in qubit space, this exchange interaction is controlled
by the state of $Q_{2}$, i.e., when $Q_{2}$ in $|0\rangle$ state, the interaction is turned
on, causing the ZZ interaction between $Q_{1}$ and $Q_{3}$, while for $Q_{2}$ in $|1\rangle$
state, the interaction is turned off.

According to Eq.$\,$(B11), Figure $\,$11 shows the calculated $J_{2(\textbf{1}),I}$
versus the anharmonicity $\alpha_{2}$ and qubit detuning $\Delta_{3}$ in the unit of
the NN coupling strength $g_{j}$ with $\omega_{1}=\omega_{3}+\alpha_{3}$. The
system parameters used are $g_{j}/2\pi=45\,\rm MHz$, and $\alpha_{j}/2\pi=-350\,\rm MHz$.
The regime indicated by the darker strip shows the working point, where the interaction
$|11\rangle_{1,3}\leftrightarrow|02\rangle_{1,3}$ is turned off, when $Q_{2}$ is in state
$|1\rangle$. From the result shown in Fig.$\,$8, one can find that in order to operate
in the dispersive regime or quasi-dispersive regime, the value of $\alpha_{2}$ should be larger
than $500\,\rm MHz$. For $\alpha_{2}$ taking a value below $500\,\rm MHz$, the dispersive model,
as well as the approximation adopted in the present work, may break down. Thus, in the
following discussion, $\alpha_{2}$ takes values larger than $500\,\rm MHz$.

Following the same procedure as that for finding the optimal working for
implementing C-iSWAP gate, we can also find the optimal working point for
realizing switchable coupling between $|11\rangle_{1,3}$ and $|02\rangle_{1,3}$,
as shown in Fig.$\,$12. Hence, it is possible to implement the
controlled-CZ gate with the switchable coupling between $|11\rangle_{1,3}$ and $|02\rangle_{1,3}$.
As shown in Figs.$\,$12(c) and 12(d), a controlled-CZ gate could be implemented in $120 \,\rm ns$.
Similar result can also be obtained for coupling between $|11\rangle_{1,3}$ and $|20\rangle_{1,3}$.
However, for the directly coupled system considered in the present work, only when the magnitudes of anharmonicity of
the two adjacent qubits are comparable to each other and have opposite signs, could the
residual parasitic ZZ coupling between adjacent qubits be heavily
suppressed \cite{R14}. Since the anharmonicity of transmon qubit is
around $200-400\,\rm MHz$, for the directly coupled system with
$\alpha_{2}$ of $665\,\rm MHz$ as shown in Fig.$\,$12,
the residual parasitic ZZ coupling between adjacent qubits cannot be heavily
suppressed, thus limiting the performance of
the controlled-CZ gate. However, for indirectly coupled system, such as a two-qubit system coupled via a bus
or a tunable coupler, the residual parasitic ZZ coupling can be heavily suppressed
without the requirement that the magnitudes of anharmonicity of the two adjacent
qubits are comparable to each other \cite{R13,R14}.

\begin{figure}[tbp]
\begin{center}
\includegraphics[width=8cm,height=6cm]{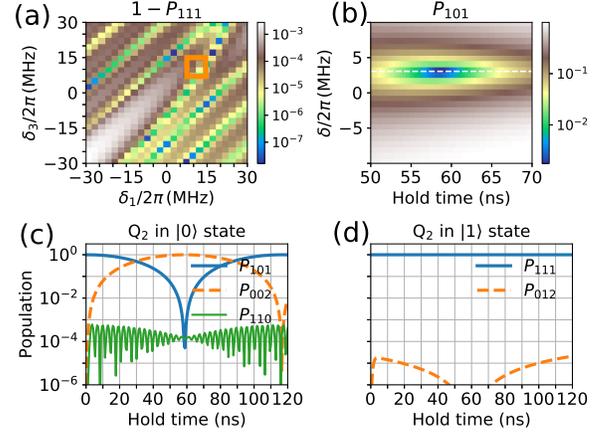}
\end{center}
\caption{Optimal working point for realizing switchable coupling between $|11\rangle_{1,3}$
and $|02\rangle_{1,3}$ with $\alpha_{2}/2\pi$ taking the value of $665\,\rm MHz$.
(a) Swap error $1-P_{111}$ versus
the frequency offsets $\delta_{1}$ and $\delta_{3}$ that are defined relative
the ideal interaction point, as shown in the inset of Fig.$\,$8. The system is
initialized in the eigenstate $|\widetilde{111}\rangle$ at the idle point, and the hold
time takes a value of 60 $\rm ns$. The (orange) square indicates the working point where the
NNN exchange interaction is turned off for $Q_{2}$ in $|1\rangle$. (b) Population $P_{101}$
versus the overshoot $\delta=\delta_{1}-\delta_{3}$ and the hold time for system initialized in the eigenstate
$|\widetilde{101}\rangle$ at the idle point. The horizontal cut (dashed line) depicts
the optimal value of overshot for enabling a full complete swap between
$|\widetilde{101}\rangle$ and $|\widetilde{002}\rangle$. (c) Population swap
$|\widetilde{101}\rangle\Leftrightarrow|\widetilde{002}\rangle$ and (d)
$|\widetilde{111}\rangle\Leftrightarrow|\widetilde{012}\rangle$ versus hold time
for system initialized in $|\widetilde{101}\rangle$ and $|\widetilde{111}\rangle$,
respectively. With optimal frequency offset and overshot obtained from (a) and (b),
the NNN exchange interaction $|11\rangle_{1,3}\leftrightarrow|02\rangle_{1,3}$ is
turned on or off depending on the state of $Q_{2}$.}
\end{figure}

\section{C-$\rm i$SWAP gate}

As mentioned in the main text, in order to implement the C-iSWAP gate in our proposed
system, the rounded trapezoid-shaped pulses are applied to adjust the frequency
of $Q_{1}$ and $Q_{3}$, thus during the gate implementation, the qubit frequencies
vary from the idle point to the interaction point, while the frequency of $Q_{2}$
keeps fixed. The rounded trapezoid-shaped pulse used in present work is described
by a time-dependent function \cite{R27}
\begin{equation}
\begin{aligned}
\omega(t)=&\omega_{i} + \frac{\omega_{I}-\omega_{i}}{2}[{\rm Erf}(\frac{t-\frac{t_{\rm ramp}}{2}}{\sqrt{2}\sigma})
-{\rm Erf}(\frac{t-t_{\rm g}+\frac{t_{\rm ramp}}{2}}{\sqrt{2}\sigma})]
\end{aligned}
\end{equation}
where $\omega_{i}$ and $\omega_{I}$ denote the idle frequency point (where the logical
states are defined as as the eigenstates of the system biased at this point, as
discussed in Appendix $\,$A) and the interaction frequency point,
respectively, the ramp time is defined as $t_{\rm ramp}=4\sqrt{2}\sigma$
with $\sigma=1\,\rm ns$, $t_{\rm g}$ represents the total time for implementing the gate
operation, and the hold time $t_{\rm hold}=t_{\rm g}-t_{\rm ramp}$ is defined
as the time-interval between the midpoints of the ramps.

\subsection{Intrinsic gate fidelity}

\begin{figure}[tbp]
\begin{center}
\includegraphics[width=8cm,height=4cm]{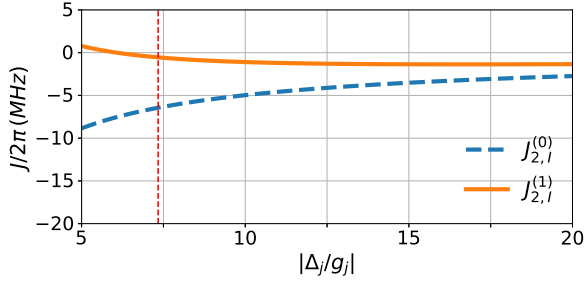}
\end{center}
\caption{Calculated strength of the effective NNN interaction
$|101\rangle\Leftrightarrow|002\rangle$ with $J_{2(\textbf{1}),I}$
and interaction $|111\rangle\Leftrightarrow|012\rangle$ with $J_{2(\textbf{0}),II}$
versus qubit detuning $\Delta_{j}$ ($\Delta_{1}=\Delta_{3}$) in the unit of the NN coupling strength $g_{j}$.
The system parameters used are $g_{j}/2\pi=45\,\rm MHz$, $\alpha_{1}/2\pi=\alpha_{3}/2\pi=-350\,\rm MHz$,
and $\alpha_{2}/2\pi=350\,\rm MH$. The vertical red dashed line indicates the working point
for the implementation of C-iSWAP gate.}
\end{figure}

To quantify the intrinsic performance of the proposed C-iSWAP gate
implementation, the metric of state-average gate fidelity is used in present
work. The fidelity is defined as \cite{R29}
\begin{equation}
\begin{aligned}
F\equiv \frac{{\rm{Tr}}(U^{\dagger}U)+|{\rm{Tr}}(U_{{\rm target}}^{\dagger}U)|^{2}}{72}
\end{aligned}
\end{equation}
where $U$ is the actual evolution operator in the logical eigenbasis at the
idle point after applying an auxiliary single-qubit $Z$ rotation on
each of the three-qubits before and after the gate implementation,
truncated to the qubit levels \cite{R27,R30,R31}, and $U_{\rm target}=U_{CXY}(\pi/2)$ is given
as Eq.$\,$(5) of the main text.

According to the system Hamiltonian in Eq.$\,$(B1), and the control pulse of Eq.$\,$(E1),
the actual evolution operator in the rotating frame with respect to $H(0)$ is
\begin{equation}
\begin{aligned}
U_{sys}=\mathcal{\hat{T}} \rm{exp}\left(-i\int_{0}^{t_{g}}H_{R}(t)dt\right),
\end{aligned}
\end{equation}
where $H_{R}(t)=e^{iH(0)t}H(t)e^{-iH(0)t}-H(0)$, and $\mathcal{\hat{T}}$
denotes the time-ordering operator. Thus, $U$ in Eq.$\,$(E2)
is given as
\begin{equation}
\begin{aligned}
U=U_{\rm{post}}\mathcal{P}U_{sys}\mathcal{P}^{\dagger}U_{\rm{pre}},
\end{aligned}
\end{equation}
where $\mathcal{P}$ is the projected operator defined in the
computational subspace of the full system, and
$U_{\rm{post}}$ and $U_{\rm{pre}}$ are
\begin{equation}
\begin{aligned}
&U_{\rm{post}}= e^{-i\phi_{1}ZII/2}e^{-i\phi_{2}IZI/2}e^{-i\phi_{3}IIZ/2},
\\&U_{\rm{pre}}= e^{-i\phi'_{1}ZII/2}e^{-i\phi'_{2}IZI/2}e^{-i\phi'_{3}IIZ/2}.
\end{aligned}
\end{equation}
Hence, the gate fidelity is obtained as
\begin{equation}
\begin{aligned}
F=\rm{maximize}_{\phi_{j},\phi'_{j}}\,\,F(\phi_{j},\phi'_{j}),
\end{aligned}
\end{equation}
taking a value of $99.97\%$ for $t_{\rm hold}=43\,\rm ns$. Aside from the control
error (as shown in Fig.$\,$4(a)) and leakage to non-qubit states (as
shown in Fig.$\,$4(b)), the residual infidelity is caused by the
coherence phase resulting from ZZ interaction between qubits, as shown in
the effective Hamiltonian of Eq.$\,$(B14). Hence, by assuming no control error
and leakage, the actual implemented unitary operator can be described by
\begin{equation}
U=\left(
\begin{array}{cccccccc}
1 & 0 & 0& 0 &0 & 0& 0 &0\\
0 & 0 & 0 & 0 & -i & 0& 0 &0\\
0 & 0 & 1 & 0& 0 & 0& 0 &0\\
0 & 0 & 0 & e^{-i\phi_{011}} & 0 & 0& 0 &0\\
0 & -i & 0 & 0& 0 & 0& 0 &0\\
0 & 0 & 0 & 0& 0 & e^{-i\phi_{101}}& 0 &0\\
0 & 0 & 0 & 0& 0 & 0& e^{-i\phi_{110}} &0\\
0 & 0 & 0 & 0& 0 & 0& 0 &e^{-i\phi_{111}}\\
\end{array}
\right),
\end{equation}
where $\phi_{s}$ ($\rm s=011,\,101,\,110,\,111$) represents the accumulated phase caused by
the ZZ interaction during the gate operation.

In Table $\rm II$ of the main text, we have also shown the accumulated phase
caused by the ZZ interaction between qubits during the gate implementation, defined as
\begin{equation}
\begin{aligned}
\phi_{s}=\rm{arg}\left(\langle s|U|s\rangle\right),
\end{aligned}
\end{equation}
i.e, the argument of the matrix element $U_{ss}=\langle s|U|s\rangle$. Strikingly,
in Table $\rm II$, one can find that the accumulated phase in state $|101\rangle$
takes the largest value $0.01728$, while in state $|111\rangle$, it is below $0.01$.
This is caused by the fact that, as shown in Fig.$\,$5, similar to the exchange interaction in qubit
space $\{|100\rangle,\,|001\rangle\}$ and $\{|110\rangle,\,|011\rangle\}$, the strength
of the virtual exchange interactions
between qubit states and non-qubit states also depends on the state of $Q_{2}$,
thus enabling the ZZ interaction (accumulated phase) to be controlled by the state
of $Q_{2}$. As shown in Fig.$\,$13, one can find that at the interaction point
(indicated as the red dashed line), the strength of the interactions between qubit
states and non-qubit states ($|101\rangle\Leftrightarrow|002\rangle$) $J_{2(\textbf{0}),I}$
is larger than that of the interaction $|111\rangle\Leftrightarrow|012\rangle$. Similar
result can also be obtained for the interaction $|101\rangle\Leftrightarrow|200\rangle$ and
the interaction $|111\rangle\Leftrightarrow|210\rangle$. This could explain the striking
feature in Table $\rm II$.

\subsection{Gate operation with the typical coupling strength}

\begin{figure}[tbp]
\begin{center}
\includegraphics[width=8cm,height=6cm]{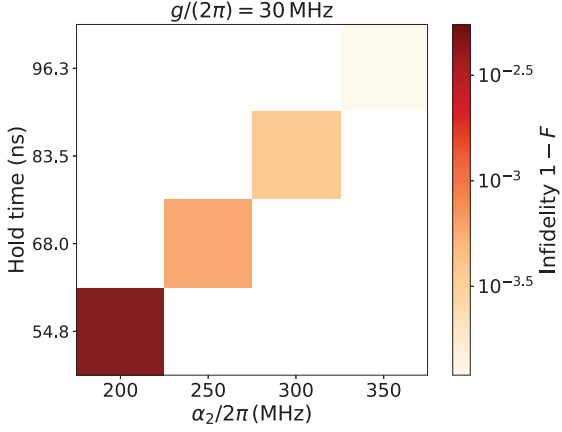}
\end{center}
\caption{The intrinsic gate fidelity versus the anharmonicity $\alpha_{2}$ and
hold time with $g_{j}/2\pi=30\,\rm MHz$. Other system parameters used here
are the same as the parameters listed in Table $\rm I$.}
\end{figure}

As mentioned in the main text, although the implementation of the native three-qubit
gates may benefit from the strong fixed coupling between adjacent qubits, this strong
coupling may make single qubit addressing \cite{R35} and the implementation of two-qubit
gates \cite{R4,R34} a challenge for system with limited frequency tunability.
However, as shown in Fig.$\,$14, with NN coupling strength of $30\,\rm MHz$ \cite{R4} and fixed ramp
time $t_{\rm ramp}=4\sqrt{2}\,\rm ns$, the intrinsic gate fidelity above $99\%$ ($99.9\%$)
can still be achieved in 50 (100) $\rm ns$. We note that choosing larger ramp time,
thus lengthening the gate time, should further reduce the leakage error \cite{R28,R33}. Therefore,
the intrinsic gate fidelity could further be improved at the expense of increased gate
time.

\subsection{Decoherence effect}

To evaluate the impact of decoherence process on the implemented gate performance,
we analyze the full system dynamics according to the Lindblad master equation
\begin{equation}
\begin{aligned}
\dot{\rho}(t)=-i[H,\rho(t)]+\sum_{l}\left(\frac{1}{T_{1}}\mathcal{C}[q_{l}]+
\frac{2}{T_{\phi}}\mathcal{C}[q_{l}^{\dag}q_{l}]\right)
\end{aligned}
\end{equation}
where $H$ is the system Hamiltonian given in Eq.$\,$(1), $\rho$ is the reduced
density matrix of the system, and
$\mathcal{C}[O]=O\rho(t)O^{\dag}-\frac{1}{2}\{O^{\dag}O,\rho(t)\}$.
In the Liouvile space \cite{R42,R43}, a ($n\times n$) density matrix $\rho$ in Hilbert
space can be mapped as a ($n^2\times 1$) vector $|\rho\rangle\rangle$, and a target unitary operator
$U_{\rm target}$ ($n\times n$) in Hilbert space can be mapped to a superoperator
$\mathcal{P}_{U_{\rm target}}=U^{\ast}_{\rm target}\otimes U_{\rm target}$ ($n^2\times n^2$).
Thus, in the Liouvile space, Eq.$\,$(E9) can be rewritten as (assuming $T_{\phi}=\infty$,
thus ignoring the dephasing term $\mathcal{L}[q_{l}^{\dag}q_{l}]$) \cite{R42,R43}
\begin{equation}
\begin{aligned}
\frac{d|\rho\rangle\rangle}{dt}=&\mathcal{L}_{t}|\rho\rangle\rangle
=\left[-i(I\otimes H-H^{T}\otimes I)\right]|\rho\rangle\rangle+
\\&\sum_{l}
\left[q_{l}^{\ast}\otimes q_{l}-\frac{1}{2}I\otimes q_{l}^{\dag}q_{l}-\frac{1}{2}q_{l}^{T}q_{l}^{\ast} \otimes I\right]|\rho\rangle\rangle,
\end{aligned}
\end{equation}
where $\ast$ denotes the complex conjugate, $T$ means the matrix transpose, $\otimes$ the Kronecker product, and $I$ is the identity operator.
Thus, the time-evolution superoperator is given as $\mathcal{P}_{U}=\mathcal{\hat{T}}{\rm{exp}}\left(\int_{0}^{t_{g}}\mathcal{L}_{t}dt\right)$.

Based on the about discussion, the average gate fidelity $F_{o}$ between target $U_{\rm target}$ and
the actually implemented operation under decoherence process is defined as \cite{R29,R39,R40,R41}
\begin{eqnarray}
\begin{aligned}
F_{o}=\frac{8(1-L_{1})+{\rm Tr}(\mathcal{P}_{U}^{\dag}\mathcal{P}_{U_{\rm{target}}})}{72},
\end{aligned}
\end{eqnarray}
with $L_{1}$ denotes the leakage of the gate operation, given as \cite{R39,R40}
\begin{eqnarray}
\begin{aligned}
L_{1}=1-\frac{1}{8}\sum_{i,j,k\in\{0,1\}}{\rm Tr}(\mathcal{P}_{U}|ijk\rangle\rangle),
\end{aligned}
\end{eqnarray}
where $|ijk\rangle\rangle$ denotes the logical qubit state $|ijk\rangle$ represented in the Liouville
space.

Similarly, we can also give the idling gate fidelity for single qubit by assuming only $T_{1}$
and $T_{\phi}$ decay process \cite{R29}, i.e,
\begin{equation}
\begin{aligned}
F_{I}&=\frac{\rm{Tr}(\sum_{k}M_{k}^{\dag}M_{k})+\sum_{k}|\rm{Tr}(M_{k})|^{2}}{6}
\\&=\frac{3+e^{-t/T_{1}}+2e^{-t(1/2T_{1}+1/T_{\phi})}}{6}
\\&\approx 1-\frac{t}{3T_{1}}-\frac{t}{3T_{\phi}},
\end{aligned}
\end{equation}
where $M_{k}$ denotes the Kraus operator describing the decoherence effect on
the qubit state, i.e., $\Lambda_{T_{1},T_{\phi}}(\rho)=\sum_{k}M_{k}\rho M_{k}^{\dag}$
with $\sum_{k}M_{k}^{\dag}M_{k}=I$. The term $\mathcal{C}[O]$ in Eq.$\,$(E9) resulting from
decoherence process can be derived from the following three Kraus operators \cite{R19,R29,R55},

\begin{equation}
M_{0}=\left(
\begin{array}{cc}
1 & 0 \\
 0 & \sqrt{1-\lambda-\gamma} \\
\end{array}
\right)
\end{equation}

\begin{equation}
M_{1}=\left(
\begin{array}{cc}
0 & 0 \\
 0 & \sqrt{\lambda} \\
\end{array}
\right),
\end{equation}

\begin{equation}
M_{2}=\left(
\begin{array}{cc}
0 & \sqrt{\gamma} \\
 0 & 0 \\
\end{array}
\right),
\end{equation}
where
\begin{equation}
\begin{aligned}
&\lambda=e^{-t/T_{1}}(1-e^{-2t/T_{\phi}}),
\\&\gamma=1-e^{-t/T_{1}}.
\end{aligned}
\end{equation}
with $T_{1}$ ($T_{\phi}$) denotes the qubit energy relaxation time (qubit dephasing time).
For our proposed three-qubit system, the idling gate fidelity can be approximated as $F=F_{I}^3$.

\end{document}